\documentclass{aa}  %esta es la version comun a dos cols.

\usepackage{placeins}  %to use FloatBarrier
\usepackage{graphicx}
\usepackage{longtable}
\usepackage{txfonts}
\begin{document}

   \title{Discovery of $\lambda$ Boo stars in open clusters}

   \titlerunning{$\lambda$ Boo stars in open clusters}
   \authorrunning{Saffe et al.}

   \author{C. Saffe\inst{1,2,5}, J. Alacoria\inst{1,5}, A. Alejo\inst{1,2,5}, A. Collado\inst{1,2,5},
           M. Flores\inst{1,2,5}, M. Jaque Arancibia\inst{3}, E. Jofré\inst{4,5},
           D. Calvo\inst{1,2,5}, P. Miquelarena\inst{1,2,5} \and E. Gonz\'alez\inst{2}
           }

\institute{Instituto de Ciencias Astron\'omicas, de la Tierra y del Espacio (ICATE-CONICET), C.C 467, 5400, San Juan, Argentina.
         \and Universidad Nacional de San Juan (UNSJ), Facultad de Ciencias Exactas, F\'isicas y Naturales (FCEFN), San Juan, Argentina.
         \and Departamento Astronom\'ia, Universidad de La Serena, av. Ra\'ul Bitr\'an 1305, La Serena, Chile.         
         \and Observatorio Astron\'omico de C\'ordoba (OAC), Laprida 854, X5000BGR C\'ordoba, Argentina
        \and Consejo Nacional de Investigaciones Cient\'ificas y T\'ecnicas (CONICET), Argentina
         }

   \date{Received xxx, xxx ; accepted xxxx, xxxx}

% \abstract{}{}{}{}{} 
% 5 {} token are mandatory
 
  \abstract
   {The origin of $\lambda$ Boo stars is currently unknown. After several efforts by many authors,
   no bona fide $\lambda$ Boo stars have been confirmed as members of open clusters.
   Their detection could provide an important test bed for a detailed study of $\lambda$ Boo stars.
   }
   {Our aim is to detect, for the first time, $\lambda$ Boo stars as members of open clusters.
   The $\lambda$ Boo class will be confirmed through a detailed abundance analysis, 
   while the cluster membership will be evaluated using a multi-criteria analysis of probable members.
   }
   {We cross-matched a homogeneous list of $\lambda$ Boo stars with a Gaia DR3 catalog of open clusters and, notably,
   we found two candidate $\lambda$ Boo stars in open clusters.
    We carried out a detailed abundance determination of the candidate $\lambda$ Boo stars and additional cluster members
   via spectral synthesis.
Stellar parameters were estimated by fitting observed spectral energy distributions (SEDs) with a grid of model atmospheres using the online
tool VOSA, Gaia DR3 parallaxes, and the PARAM 1.3 interface. Then, the abundances were determined iteratively
for 22 different species by fitting synthetic spectra using the SYNTHE program together with local thermodynamic equilibrium
(LTE) ATLAS12 model atmospheres. The abundances of the light elements C and O were corrected by non-LTE effects.
The complete chemical patterns of the stars were then compared to those of $\lambda$ Boo stars.
We also performed an independent cluster membership study using Gaia photometry and radial velocities
with a multi-criteria analysis.
}
   {
 For the first time, we present the surprising finding of two $\lambda$ Boo stars as members of open clusters:
 HD 28548 belongs to the cluster HSC 1640 and
 HD 36726 belongs to the cluster Theia 139.
 This was confirmed using a detailed abundance analysis,
 while the cluster membership was independently analyzed using Gaia DR3 data and radial velocities.
We compared the $\lambda$ Boo star HD 36726 with other cluster members
and showed that the $\lambda$ Boo star was originally born with a near-solar composition.
This also implies one of the highest chemical differences detected between two cluster members ($\sim$0.5 dex).
In addition, we suggest that the $\lambda$ Boo peculiarity strongly depletes heavier metals,
but could also slightly modify lighter abundances such as C and O.
We also found that both $\lambda$ Boo stars belong to the periphery of their respective clusters. 
This would suggest that $\lambda$ Boo stars avoid the strong photoevaporation
by UV radiation from massive stars in the central regions of the cluster.
We preliminarily suggest that peripheral location appears to be a necessary, though not sufficient, 
condition for the development of $\lambda$ Boo peculiarity.
We also obtained a precise age determination for the $\lambda$ Boo stars
HD 28548 (26.3$\pm$1.4 Myr) and HD 36726 (33.1$\pm$1.1 Myr), being
one of the most precise age determinations of $\lambda$ Boo stars.
We strongly encourage analyzing additional cluster members,
which could provide important insights for studying the origin of $\lambda$ Boo stars.
   }
   {We have confirmed, for the first time, that two $\lambda$ Boo stars belong to open clusters.
This remarkable finding could make open clusters excellent laboratories for studying the origin of $\lambda$ Boo stars.
   }
   

   \maketitle
%
%________________________________________________________________

\section{Introduction}

For many years, the nature of the $\lambda$ Boo stars has been controversial.
They are a class of metal-poor Population I A-type stars first discovered in the work of \citet{morgan43}.
The $\lambda$ Boo stars are markedly metal deficient (up to $\sim$2 dex in the most extreme cases), 
but have nearly solar abundances of the lighter elements CNO and S
\citep[e.g. ][]{kamp01,andrievsky02,heiter02,alacoria22}.
Most $\lambda$ Boo stars have been discovered through spectral classification
\citep[see, e.g. ][]{murphy15,gray17,murphy20},
although a detailed abundance determination has been suggested 
to confirm their membership to the class \citep[e.g. ][]{andrievsky02,heiter02,murphy15,gray17,alacoria22,alacoria25}.
These rare stars \citep[about $\sim$2\% of the A-type stars, ][]{gray-corbally98,paunzen01b} 
present a considerable challenge to our understanding of stellar atmospheres.

The detection of $\lambda$ Boo stars as members of multiple systems is important for several reasons.
For instance, this could help to precisely determine their evolutionary state,
which motivated searches of $\lambda$ Boo stars in open clusters by several authors
\citep[e.g. ][]{paunzen-gray97,gray-corbally98,paunzen01a,paunzen01b,gray-corbally02,paunzen03b,paunzen14}.
Furthermore, the composition of late-type stars in a multiple system could be used as a proxy of the initial
composition from which $\lambda$ Boo stars were born \citep[e.g. ][]{alacoria22,alacoria25}.
In addition, the comparison of a $\lambda$ Boo star with other early-type stars in the system
could be used to test formation models such as the interaction with a diffuse cloud
\citep[see, e.g. ][]{paunzen12a,paunzen12b,alacoria22}.
The works mentioned assume that multiple systems are a single population of co-evolutionary stars,
that is, the stars formed together from the same molecular cloud.
In this way, the detection of $\lambda$ Boo stars in multiple systems is considered a valuable discovery,
providing a key laboratory to study their origin in detail.

However, detecting $\lambda$ Boo stars as members of multiple systems has proven to be a difficult task.
For the case of binary systems, \citet{paunzen12a,paunzen12b} identified a $\sim$dozen of
$\lambda$ Boo stars as members of early-type binary systems,
while \citet{alacoria25} recently identified new systems including late-type companions.
On the other hand, to our knowledge no bona fide $\lambda$ Boo star is known
as a member of an open cluster.
In order to determine the evolutionary state of $\lambda$ Boo stars, 
extensive searches were performed in 32 different open clusters 
\citep[see, e.g. ][]{paunzen-gray97,gray-corbally98,paunzen01a,paunzen01b,gray-corbally02,paunzen03b,paunzen14}.
The searches mentioned mainly included intermediate-age open clusters, that is, with ages between 10$^{7}$ yr and 10$^{9}$ yr, approximately.
Notably, although several chemically peculiar Am and Ap stars were detected
in the same clusters \citep[e.g. ][]{gray-corbally02,paunzen14}, no $\lambda$ Boo stars were identified.
For instance, \citet{paunzen-gray97}  and \citet{paunzen01a} reported a candidate $\lambda$ Boo star in the open cluster NGC 2264
(HD 261904 $=$ NGC 2264 \# 138) and three candidates in the periphery of the Orion OB1 association. 
However, \citet{andrievsky02} determined suprasolar abundances for Mg and Fe in HD 261904,
while \citet{murphy15} classified this object as an uncertain member of the $\lambda$ Boo class.
We also note that the Orion OB1 association presents subgroups of stars (OB1a, OB1b, OB1c and OB1d) with 
different ages and dynamical properties, rather than a single population
\citep[see, e.g. ][ and references therein]{bally08}.
The examples mentioned illustrate that the detection of $\lambda$ Boo stars in multiple systems
is a challenging task and deserves to be further explored.

With the aim of providing a consistent and homogeneous membership of $\lambda$ Boo stars,
a number of works reevaluated every $\lambda$ Boo star previously
reported in the literature \citep[see ][]{murphy15,gray17,murphy20}.
Together, these three works comprise a complete and homogeneous sample of 118 (predominantly southern) $\lambda$ Boo stars.
On the other hand, \citet{hunt-reffert24} performed a blind all-sky search for open clusters
using Gaia DR3 data of 729 million stars down to magnitude G$\sim$20, 
and recovered 5647 bound open clusters (1441 of which are new detections).
Then, with the aim of detecting $\lambda$ Boo stars in multiple systems, 
we cross-matched the list of 118 $\lambda$ Boo stars with this catalog of clusters.
Surprisingly, we found two $\lambda$ Boo stars with a very high membership probability:
HD 28548 member of HSC 1640 (Prob$\sim$95\%, log t$\sim$7.2) and 
HD 36726 member of Theia 139 (Prob$\sim$100\%, log t$\sim$8.0).
In particular, HD 36726 is one of the three $\lambda$ Boo stars reported by \citet{paunzen-gray97}
as member of the Orion OB1 association.
We note that HSC 1640 and Theia 139 are classified as open clusters in different works \citep{hunt-reffert23,cavallo24},
while \citet{hunt-reffert24} caution that both clusters are possibly dissolving
(we will return to this point in the discussion).
The remarkable finding of these two $\lambda$ Boo stars encourages
us to study them as well as other members of their multiple systems.
This could provide the possibility of having a valuable test bed for a detailed study
of $\lambda$ Boo stars.

In this work, we present a detailed chemical analysis of two
candidate $\lambda$ Boo stars together with other objects that belong to the same multiple system.
This would require 
1) to confirm the $\lambda$ Boo class of the two candidates with a detailed chemical analysis, and
2) to verify that the stars in the multiple system belong to a single population through a membership analysis.
We analyzed the chemical composition of both stars together with additional cluster members, 
with an initial membership suggested by the work of \citet{hunt-reffert24}.
We also performed a independent membership study using a multi-criteria analysis,
showing that both candidate $\lambda$ Boo stars are members of their multiple systems.
However, some of the additional stars analyzed should be taken with caution.
Studying $\lambda$ Boo stars along with their stellar 
siblings\footnote{In this work, we call "siblings" to those stars that belong to the same stellar population.}
could provide important insights about the origin of $\lambda$ Boo stars.

This work is organized as follows. 
In Sect. 2, we describe the sample and observations.
In Sect. 3, we present the stellar parameters and chemical abundance analysis. 
In Sect. 4, we present our independent cluster membership analysis, while
in Sect. 5, we discuss the results.
Finally, in Sect. 6, we highlight our main conclusions.

\section{Sample and observations} \label{observaciones}

The sample of stars used in this work was derived as follows.
The candidate $\lambda$ Boo stars HD 28548 and HD 36726 were obtained by
cross-matching the homogeneous list of 118 $\lambda$ Boo stars with the
cluster catalog of \citet{hunt-reffert24}, as previously explained.
Then, additional cluster members were initially suggested by selecting relatively bright stars
with high membership probability ($>$ 90\%) in the same cluster catalog \citep{hunt-reffert24}.
We derived a detailed chemical composition for all these stars.
Then, we performed a independent membership analysis including radial velocities
and confirmed that 
both candidate $\lambda$ Boo stars are members of the clusters,
while some of the additional stars do not appear to be true cluster members.
Our cluster membership analysis including an age derivation for the clusters,
is presented in section \ref{section.membership}.

We present in Table \ref{table.parallax} the visual magnitude V, coordinates, proper motions,
parallax and signal-to-noise per pixel (@5500 \AA), for the stars studied in this work.
The spectral data for the stars in this work were acquired through the
Gemini High-resolution Optical SpecTrograph (GHOST),
which is attached to the 8.1 m Gemini South telescope at Cerro Pach\'on, Chile.
GHOST is illuminated via 1.2" integral field units that provide the
input light apertures. The spectral coverage of GHOST between
360-900 nm is appropriate for deriving stellar parameters and
chemical abundances using several features. 
It provides a high resolving power R$\sim$50000 in the standard resolution 
mode\footnote{https://www.gemini.edu/instrumentation/ghost}.
The read mode was set to medium, as recommended for relatively bright targets.
The observations were taken between October 4, 2024 and October 26, 2024 (PI: Carlos Saffe, Program ID: GS-2024B-FT-210)
under a Fast Turnaround (FT) observing mode\footnote{https://www.gemini.edu/observing/phase-i/ft},
using the same spectrograph configuration for all stars.
The exposure times varied between 40 sec and 280 sec depending on the star,
obtaining an average signal-to-noise ratio (S/N) $\sim$315 per pixel measured near $\sim$5500 {\AA}.
The spectra were reduced using the GHOST data reduction pipeline v1.1.0, which works under 
DRAGONS\footnote{https://www.gemini.edu/observing/phase-iii/reducing-data/dragons-data-reduction-software}. This is
a platform for the reduction and processing of astronomical data.

\begin{table*}
\centering
\caption{Magnitudes and astrometric data for the stars analyzed in this work.}
\begin{tabular}{lccccccc}
\hline
Star        & V      & $\alpha$    & $\delta$     & $\mu_{\alpha}$ & $\mu_{\delta}$ & $\pi$   & S/N  \\
            &        & J2000       & J2000        & [mas/yr]       & [mas/yr]       & [mas]   & @5500 \AA    \\
\hline
HD 28548    &  9.22  & 04 29 27.25 & -15 01 51.11 &  2.495         & -3.701         &  4.2647 & 350 \\  
HD 25674    &  8.69  & 04 04 01.82 & -11 44 54.17 &  2.024         & -5.578         &  5.0265 & 320 \\  
BD-06 984   & 10.43  & 04 45 14.91 & -06 39 06.94 &  3.037         & -2.074         &  3.9520 & 350 \\  
BD-08 924   &  9.76  & 04 43 26.32 & -08 30 53.88 & -0.971         & -3.464         &  3.9709 & 310 \\  
BD-12 905   & 10.23  & 04 28 36.89 & -12 09 06.36 &  2.843         & -3.377         &  4.1881 & 280 \\  
\hline
HD 36726    &  8.82  & 05 33 51.73 & -00 04 36.34 & -2.940         & -4.521         &  3.1016 & 330 \\  
HD 37333    &  8.51  & 05 37 40.47 & -02 26 36.84 & -2.686         & -3.494         &  2.9822 & 240 \\  
HD 37187    &  8.15  & 05 36 37.07 & -01 01 40.77 & -3.081         & -3.896         &  2.8306 & 340 \\  
HD 290541   & 10.07  & 05 30 59.84 & -02 02 02.48 & -2.739         & -4.352         &  3.1279 & 350 \\  
HD 290621   & 10.65  & 05 34 21.90 & +00 40 56.67 & -2.715         & -4.385         &  2.9754 & 290 \\  
\hline
\end{tabular}
\tablefoot{The two groups correspond (in principle) to the clusters HSC 1640 and Theia 139.
However, some of the stars then resulted non-members (see section \ref{section.membership}).}
\label{table.parallax}
\end{table*}

\section{Stellar parameters and abundance analysis} 

\begin{figure*}
\centering
\includegraphics[width=8.0cm]{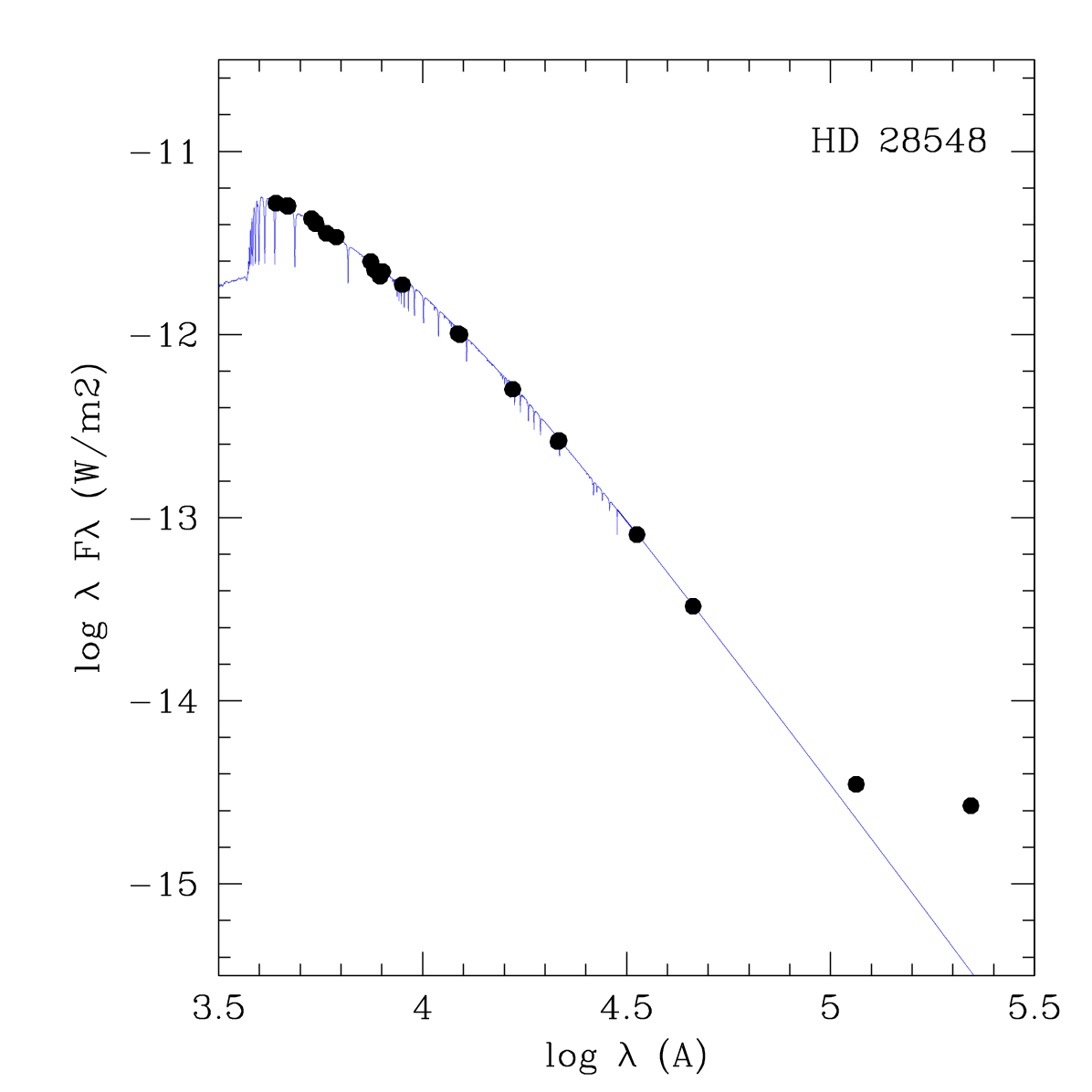}
\includegraphics[width=8.0cm]{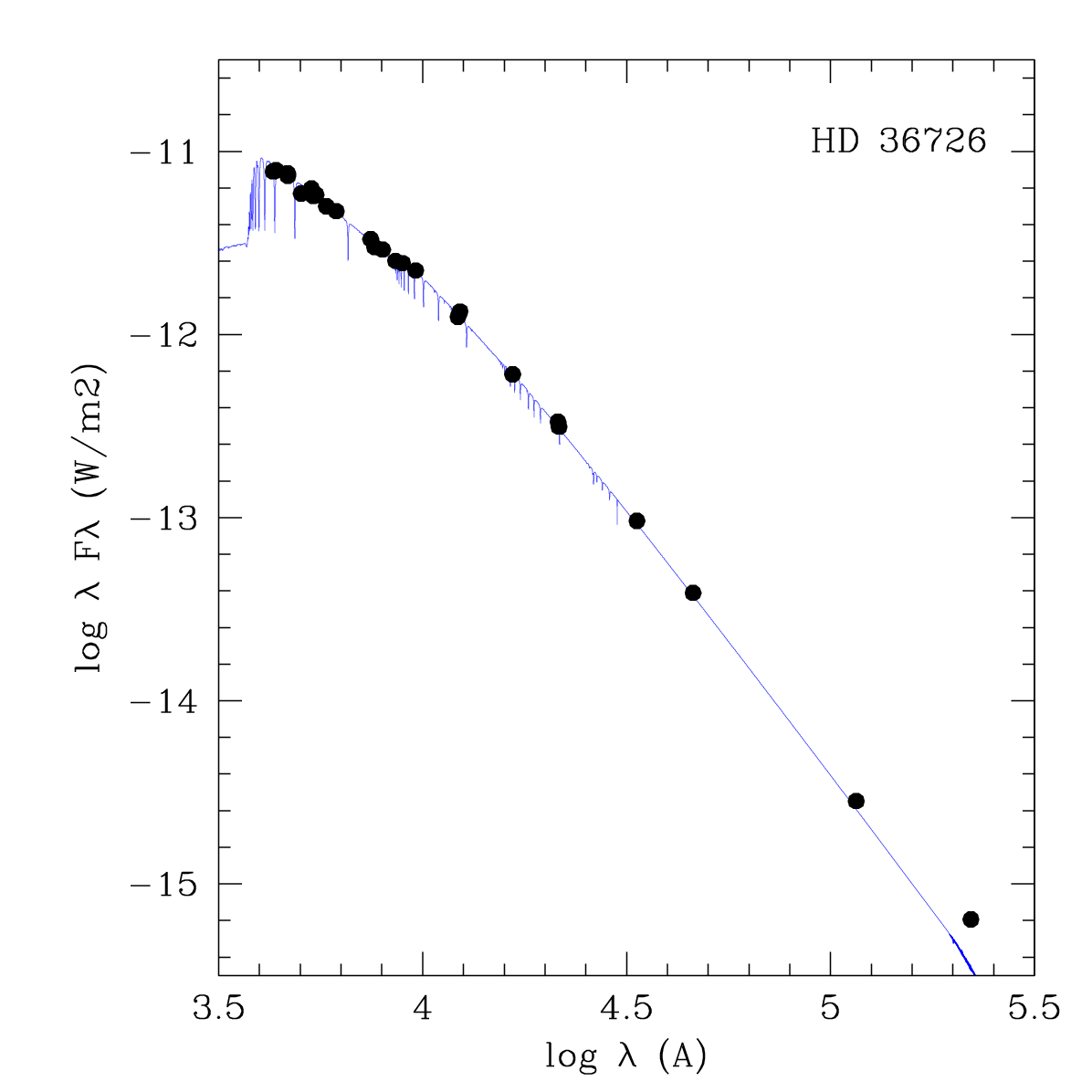}
\caption{Spectral energy distribution (black circles) and synthetic spectra (blue continuous line)
corresponding to the stars HD 28548 and HD 36726 (left and right panels).}
\label{fig.seds}
\end{figure*}

We estimated the effective temperatures (T$_\mathrm{eff}$)
for the stars in our sample by using the 
Virtual Observatory Sed Analyzer\footnote{http://svo2.cab.inta-csic.es/theory/vosa/} \citep[VOSA, ][]{bayo08}.
This tool allows fitting spectral energy distributions (SEDs) constructed from photometric data
using different atmospheric models.
Observed SEDs were unreddened by VOSA using the extinction maps of 
\citet{schlegel98} and following the procedure of \citet{bilir08} to derive A$_{v}$.
We used a grid of Kurucz-NEWODF models \citep{kurucz93} covering T$_\mathrm{eff}$ between 3500 K - 13000 K,
with a step of 250 K.
We present an example of observed and synthetic SEDs for the stars HD 28548 and HD 36726 in Fig. \ref{fig.seds}.
Then, we performed a Bayesian estimation of surface gravities $\log g$ using Gaia DR3 parallaxes with
the PARAM 1.3 interface\footnote{http://stev.oapd.inaf.it/cgi-bin/param$\_$1.3} \citep{dasilva06}.
Temperatures and gravities derived for the stars in our sample are presented in Table \ref{table.params}.

In the next step, we estimated projected rotational velocities $v\sin i$ 
by fitting the line \ion{Mg}{II} 4481.23 \AA\
and then refined by fitting most \ion{Fe}{I} and \ion{Fe}{II} lines in the spectra,
similar to previous works \citep{saffe-levato14,saffe20,saffe21}.
Synthetic spectra were calculated using the program SYNTHE \citep{kurucz-avrett81} together with
ATLAS12 \citep{kurucz93} model atmospheres.
These spectra are then convolved with a rotational profile 
and with an instrumental profile for the GHOST spectrograph (resolving power R $\sim$ 50000).
The resulting $v\sin i$ values are shown in the 6th column of Table \ref{table.params},
covering between 22.4 $\pm$ 1.6 km s$^{-1}$ and 129.6 $\pm$ 2.7 km s$^{-1}$ for the stars in our sample.
Microturbulence velocity (v$_\mathrm{micro}$) was estimated as a function of T$_{\rm eff}$ following the
formula of \citet{gebran14}, which is valid for $\sim$6000 K $<$ T$_{\rm eff}$ $<$ $\sim$10000 K.
We adopted an uncertainty of $\sim$25 $\%$  for v$_\mathrm{micro}$, as suggested by \citet{gebran14},
and then this uncertainty was taken into account in the abundance error calculation.

\begin{table*}
\centering
\caption{Fundamental parameters derived for the stars in this work.}
\begin{tabular}{lccccr}
\hline
Star            &  T$_{\rm eff}$  & $\log g$        & [Fe/H]           & v$_\mathrm{micro}$ & $v\sin i$     \\
                &   [K]           &  [dex]          & [dex]            & [km s$^{-1}$]      & [km s$^{-1}$] \\
\hline
HD 28548        &  8500 $\pm$ 250 & 4.27 $\pm$ 0.04 & -1.21 $\pm$ 0.16 & 3.15 $\pm$ 0.79 &  85.0 $\pm$ 0.8 \\ 
HD 25674        &  8850 $\pm$ 250 & 4.25 $\pm$ 0.06 & -0.07 $\pm$ 0.21 & 2.84 $\pm$ 0.71 & 128.3 $\pm$ 2.2 \\ 
BD-06 984       &  7000 $\pm$ 250 & 4.26 $\pm$ 0.04 & -0.06 $\pm$ 0.22 & 2.30 $\pm$ 0.58 &  93.3 $\pm$ 1.5 \\ 
BD-08 924       &  7750 $\pm$ 250 & 4.23 $\pm$ 0.04 & +0.37 $\pm$ 0.13 & 2.00 $\pm$ 0.50 &  54.0 $\pm$ 1.0 \\ 
BD-12 905       &  6750 $\pm$ 250 & 4.23 $\pm$ 0.06 & -0.04 $\pm$ 0.20 & 1.86 $\pm$ 0.47 &  22.4 $\pm$ 1.6 \\ 
HD 36726        &  9000 $\pm$ 250 & 4.13 $\pm$ 0.04 & -0.72 $\pm$ 0.20 & 2.67 $\pm$ 0.67 &  99.6 $\pm$ 0.9 \\ 
HD 37333        &  9500 $\pm$ 250 & 3.96 $\pm$ 0.05 & -0.08 $\pm$ 0.20 & 2.05 $\pm$ 0.51 &  43.6 $\pm$ 0.7 \\ 
HD 37187        & 10500 $\pm$ 250 & 4.02 $\pm$ 0.04 & -0.09 $\pm$ 0.15 & 0.95 $\pm$ 0.24 & 129.6 $\pm$ 2.7 \\ 
HD 290541       &  7750 $\pm$ 250 & 4.24 $\pm$ 0.05 & -0.18 $\pm$ 0.16 & 3.21 $\pm$ 0.80 &  83.0 $\pm$ 1.9 \\ 
HD 290621       &  7000 $\pm$ 250 & 4.19 $\pm$ 0.05 & -0.12 $\pm$ 0.18 & 2.30 $\pm$ 0.58 & 106.4 $\pm$ 1.4 \\ 
\hline
\end{tabular}
\normalsize
\label{table.params}
\end{table*}

Chemical abundances were determined by applying an iterative procedure, similar to previous works
\citep[e.g. ][]{saffe20,saffe21,saffe22,alacoria22,alacoria25}.
We start the process by initially adopting solar abundances from \citet{asplund09} and 
deriving an ATLAS12 \citep{kurucz93} model atmosphere.
The corresponding abundances were then obtained by fitting
the observed spectra with the program SYNTHE \citep{kurucz-avrett81}.
With the new abundance values, we derived a new model atmosphere and restarted the process again.
In each step, opacities were calculated for an arbitrary composition and v$_\mathrm{micro}$ using the opacity
sampling (OS) method. 
In this way, parameters and abundances were consistently derived
using specific opacities rather than solar-scaled values,
until reach the same input and output abundance values \citep[for more details, see ][]{saffe21}.
Possible differences originating from the use of solar-scaled opacities instead of an arbitrary
composition were recently estimated for the case of solar-type stars \citep{saffe18,saffe19}.

\begin{figure}
\centering
\includegraphics[width=8cm]{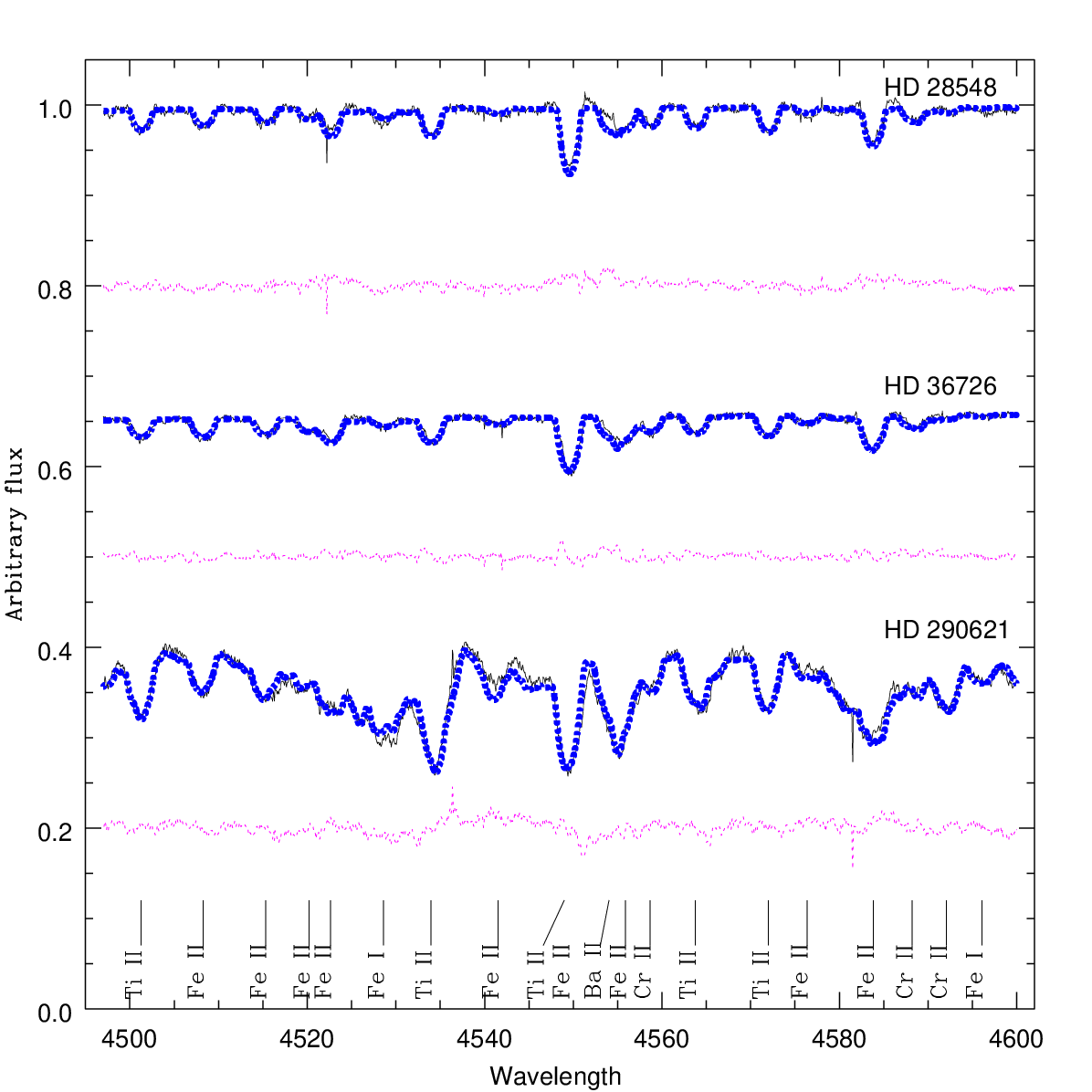}
\caption{Observed, synthetic, and difference spectra (black, blue dotted, and magenta lines) 
for some stars in our sample.}
\label{fig.region2}
\end{figure}

We derived the chemical abundances for 22 different species, including 
\ion{Li}{I}, \ion{C}{I}, \ion{O}{I}, \ion{Na}{I}, \ion{Mg}{I}, \ion{Mg}{II},
\ion{Al}{I}, \ion{Si}{II}, \ion{Ca}{I}, \ion{Ca}{II}, \ion{Sc}{II},
\ion{Ti}{II}, \ion{Cr}{II}, \ion{Mn}{I},
\ion{Fe}{I}, \ion{Fe}{II}, \ion{Co}{I}, \ion{Ni}{II},
\ion{Zn}{I}, \ion{Sr}{II}, \ion{Y}{II}, and \ion{Ba}{II}.
The atomic line list and laboratory data used in this work are the same described in 
previous works of our team \citep{saffe21,saffe22,alacoria22,alacoria25}.
Figure \ref{fig.region2} presents an example of observed, synthetic, and difference spectra 
(black, blue dotted, and magenta lines) for some stars in our sample.
There is a good agreement between modeling results and observations for the lines of different chemical species.

For each individual specie, we estimated the uncertainty in the corresponding abundance value
by including different sources.
The measurement error, e$_{1}$, was estimated from the line-to-line dispersion
as $\sigma/\sqrt{n}$, where $\sigma$ is the standard deviation and n is the number of lines.
For elements with only one line, we adopted for $\sigma$ the standard deviation of the iron lines.
Then, we determined the contribution to the abundance error due to the uncertainty 
in stellar parameters. We modified T$_{\rm eff}$, $\log g$, and v$_\mathrm{micro}$ by their
uncertainties and recalculated the abundances, obtaining the corresponding
differences e$_{2}$, e$_{3}$, and e$_{4}$ (we adopt a minimum of 0.01 dex for these errors).
Finally, the total error e$_{tot}$ was estimated as the quadratic sum of e$_{1}$, e$_{2}$, e$_{3}$, and e$_{4}$.
We discuss the chemical pattern for each star in Appendix \ref{section.abundances}, 
comparing them with an average pattern of $\lambda$ Boo stars and other chemically peculiar stars.
The average pattern of $\lambda$ Boo stars is the same used in \citet{alacoria22}:
basically, we adopted the data derived by \citet{heiter02}, but excluding those stars
without CNO values.
The abundances with their total error e$_{tot}$, the individual errors e$_{1}$ to e$_{4}$, and the
number of lines n, are presented in Tables \ref{tab.abunds.HD28548} to \ref{tab.abunds.HD290621} of the Appendix \ref{section.tables}.

\subsection{NLTE effects}

In the upper main-sequence, a parameter space occupied by $\lambda$ Boo stars,
non-local thermodynamic equilibrium (NLTE) abundance corrections
are relevant for light-elements \citep[see, e.g. ][]{paunzen99,alacoria22,alacoria25}.
For instance, an average \ion{O}{I} correction of -0.5 dex was derived by \citet{paunzen99}
for a sample of $\lambda$ Boo stars; while for \ion{C}{I}, they estimated an average 
correction of -0.1 dex.
\citet{rentzsch96} derived neutral carbon NLTE abundance corrections by using a multilevel 
model atom for stars with T$_\mathrm{eff}$ between 7000 K and 12000 K, log g between
3.5 and 4.5 dex, and metallicity from -0.5 dex to +1.0 dex.
She showed that \ion{C}{I} NLTE effects are negative (calculated as NLTE-LTE) and
depend primarily on equivalent width W$_{eq}$.
Near $\sim$7000 K, the three lower levels of \ion{C}{I} are always in LTE; however,
increasing the T$_\mathrm{eff}$ values increases the underpopulation of these levels
respect to LTE by UV photoionization.
Thus, we estimated NLTE abundance corrections of \ion{C}{I} for the stars
in our sample by interpolating in their Figs. 7 and 8 as a function of 
T$_\mathrm{eff}$, W$_{eq}$, and metallicity. We applied a similar correction in 
previous works \citep{alacoria22,alacoria25},
which allows the comparison of abundance values.

For the case of \ion{O}{I}, we performed NLTE abundance corrections following the
work of \citet{sitnova13},
who determined LTE and NLTE abundances using a model atom with 51 levels.
\citet{sitnova13} showed that NLTE effects lead to an strengthening of \ion{O}{I} lines,
producing a negative NLTE correction.
They calculated NLTE corrections for a grid of model atmospheres, including stars
with spectral types from A to K (T$_\mathrm{eff}$ between 10000 and 5000 K).
Then, we estimated NLTE abundance corrections of \ion{O}{I} (IR triplet 7771 \AA)
for the stars in this work, interpolating based on Table 11 of \citet{sitnova13}, as
a function of T$_\mathrm{eff}$. We note that other \ion{O}{I} lines present corrections 
lower than $\sim$-0.02 dex \citep[see, e.g., Table 5 of ][]{sitnova13}.

\section{Membership to open clusters}  \label{section.membership}

In this section, we analyze the membership of all stars in our sample
to the open clusters HSC~1640 and Theia~139,
including the $\lambda$~Boo stars HD~28548 and HD~36726.
We applied a multi-criteria analysis to determine the cluster membership
using a kinematic and photometric selection of probable members,
following a procedure similar to previous works \citep[e.g. ][]{alejo20}.
We used the initial membership suggested by \citet{hunt-reffert24} for each cluster
\citep[using astrometric and photometric Gaia DR3 data, ][]{gaiadr3},
but also complemented with radial velocities available in the literature and measured in our spectra
(see Table \ref{vr} of Appendix C).

First, we determined a kinematic membership by comparing individual radial velocities with the cluster mean,
adopting a criteria based on their uncertainty.
We measured the radial velocities of all observed stars by cross-correlation
using the {\sc IRAF} task \emph{fxcor} with the GHOST spectra.
In the selection of spectral regions, we excluded hydrogen lines, interstellar lines, and regions lacking spectral lines. 
The template spectra for the cross-correlations were selected from the database of \citet{bertone08}. 
The synthetic templates were then convolved with appropriate rotational profiles. 
The relative velocities between objects and templates were measured by fitting a Gaussian to the correlation peak. 
We also took into account the radial velocities obtained from the surveys RAVE DR6 \citep{rave},
Gaia DR3 \citep{gaiadr3} and APOGEE \citep{apogee}. 
Gaia DR3 velocities were corrected as proposed by \citet{katz23}, for those stars where the integrated G$_{rvs}$ magnitude is
G$_{rvs}$ $\geq$ 11 mag, and for stars where 6 mag $\leq$ G$_{rvs}$ $\leq$ 12 mag, according to \citet{blomme23}. % xxx (redaccion) 
Because we have few spectra and few radial velocity measurements, we do not rule out that some objects 
may present spectroscopic variability.
For those stars with more than one radial velocity measurement, 
we calculated the average velocity ($\overline{V}$) and its uncertainty ($\varepsilon_{\overline{V}}$)
using the expressions of \citet{gonzalez00}. 
In those equations, the error is calculated considering both the individual errors and the dispersion of the measurements. 
The results of these calculations are listed in Table \ref{vr}.
In particular, we note that the star HD~290572 in Theia~139 is a spectroscopic binary \citep{gaia22},
so we used its center-of-mass velocity in the calculations.

\begin{table}
 \caption{Average radial velocities and membership for the stars in our sample.} \label{vr.old}
 \begin{center}
 \begin{tabular}{lcccc}
  \hline\hline
  Star      &      $\overline{V}$   &    $\varepsilon_{\overline{V}}$  &     n     &     Cluster \\  
            &        [km s$^{-1}$]  &         [km s$^{-1}$]           &           & \\
  \hline
  HD~28548  &         16.24         &               3.41            &     1     &       HSC~1640\\
  HD~25674  &         10.73         &               2.36            &     2     &       HSC~1640?\\
  BD-06~984 &         18.55         &               4.32            &     2     &       NM\\
  BD-08~924 &         21.15         &               1.48            &     2     &       HSC~1640?\\
  BD-12~905 &         14.21         &               0.56            &     2     &       NM    \\
  HD~36726  &         17.89         &               3.69            &     2     &       Theia~139\\
  HD~37333  &         29.13         &               1.11            &     2     &       NM\\
  HD~37187  &         16.69         &               5.53            &     1     &       Theia~139\\
  HD~290541 &         13.74         &               0.76            &     3     &       NM\\
  HD~290621 &         15.23         &               1.43            &     3     &       Theia~139\\
  
  \hline
 \end{tabular} 
 \end{center}
 \end{table}

To calculate the cluster mean radial velocity and determine the kinematic membership, 
we analyzed the radial velocity distribution of the observed stars, 
those available in the bibliography and probable members according to \citet{hunt-reffert24}.
First, we fitted the radial velocity distribution with the Levenberg–Marquardt least-squares optimization technique
from the SciPy library in PYTHON, and obtained a first mean value for each cluster. 
We used a bin size of 1~km s$^{-1}$ in the calculation. 
Taking into account this mean cluster velocity, the radial velocity of the stars and theirs uncertainties, 
we adopted a selection criteria based on uncertainty. 
We selected as probable radial velocity members those stars whose radial velocity difference
with the cluster mean is smaller than its uncertainty,
a criteria previously adopted by \citet{alejo20}.
In the final calculation of the radial velocity of the cluster we included all stars classified as members. 
Since radial velocity uncertainty could vary from star to star, in the cluster mean we applied weights proportional to 
$(\epsilon^2 + \Delta^2)^{-1}$, where $\epsilon$ is the uncertainty of the star radial velocity and $\Delta=1$ km~s$^{-1}$. 
The reason for including the constant $\Delta$ is that the radial velocity of an individual star, 
as an estimator of the mean cluster velocity, has an uncertainty of at least the internal velocity of the cluster,
which in open clusters is typically on the order of 1 km~s$^{-1}$. 
The final values of the mean radial velocity of the clusters are 
$13.2 \pm 0.4$ km~s$^{-1}$ for HSC~1640 and 
$16.3 \pm 0.3$ km~s$^{-1}$ for Theia~139. 
In this way, we found that the $\lambda$~Boo stars HD 28548 and HD 36726 are probable radial velocity members.

\begin{figure*}
\begin{center}
\centering
\includegraphics[width=8cm]{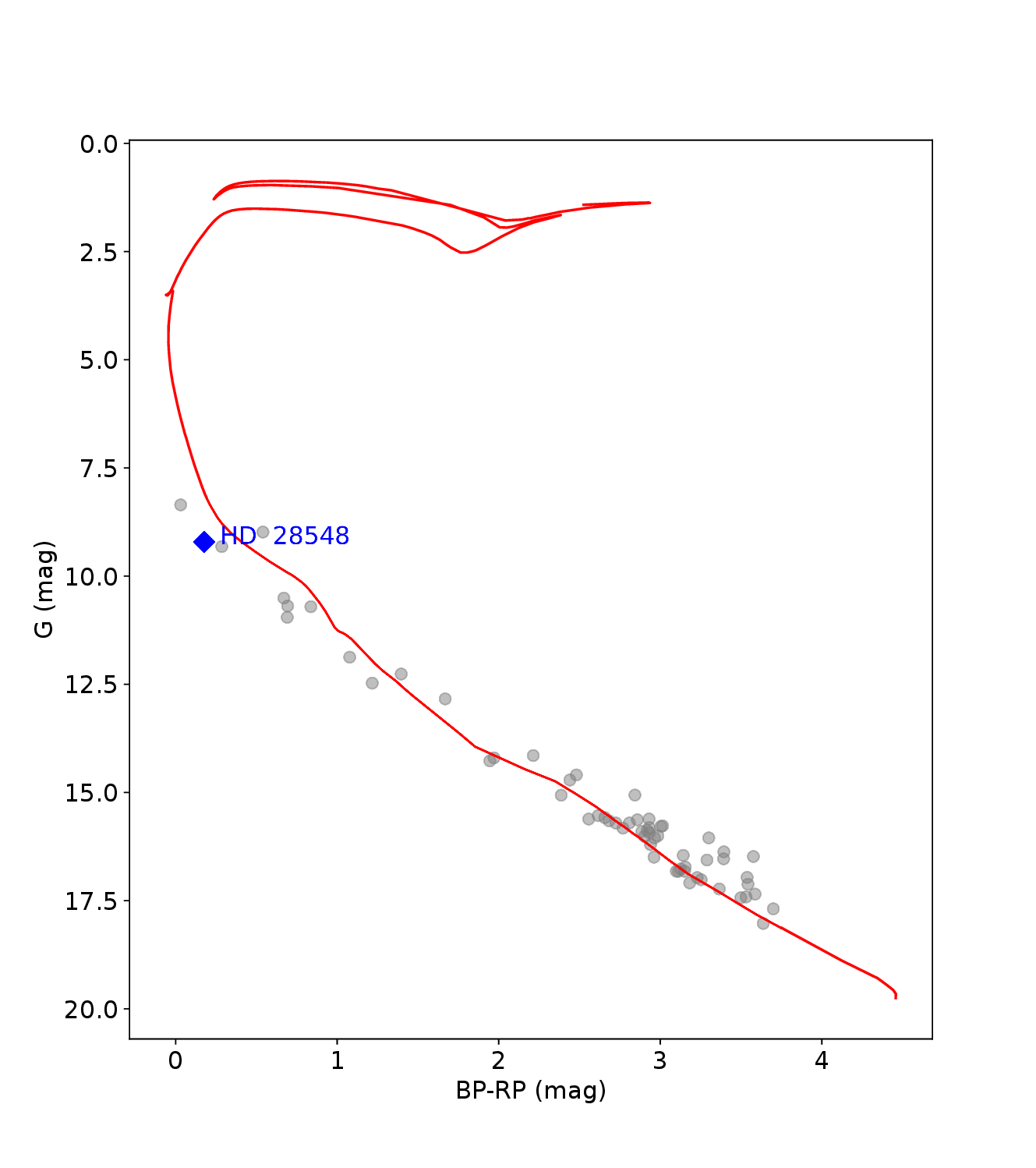}
\includegraphics[width=8cm]{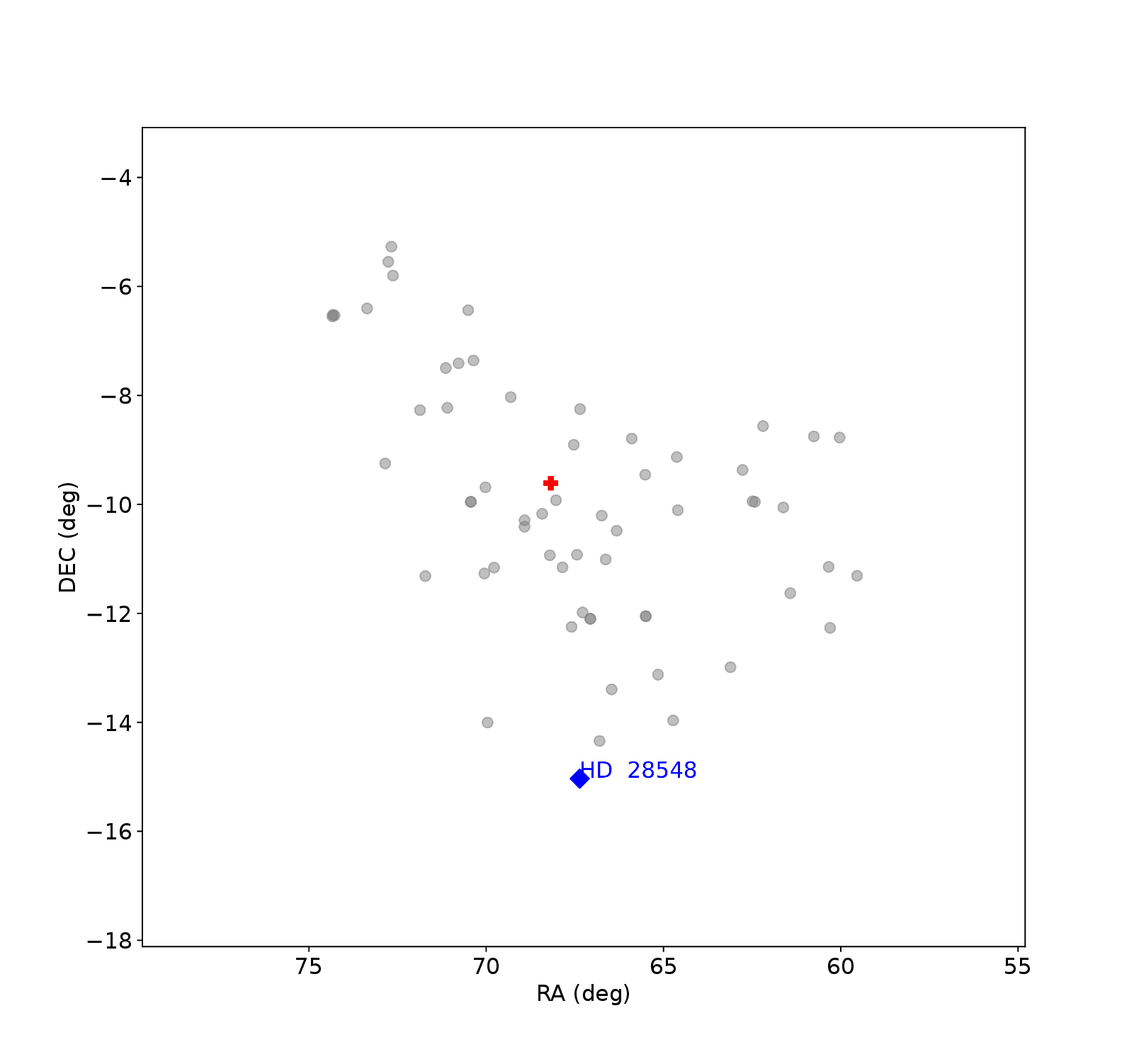}
\includegraphics[width=8cm]{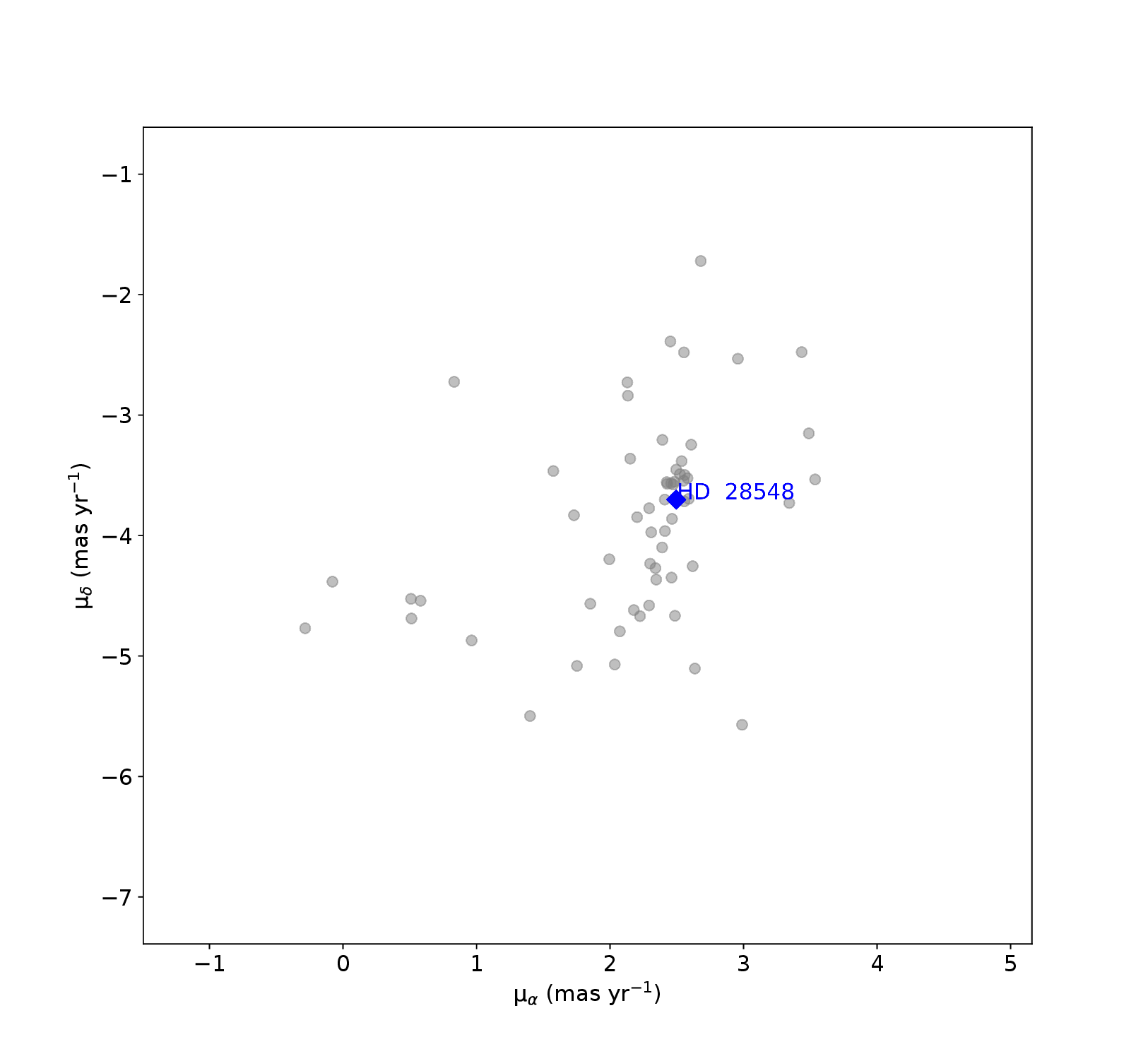} 
\includegraphics[width=8cm]{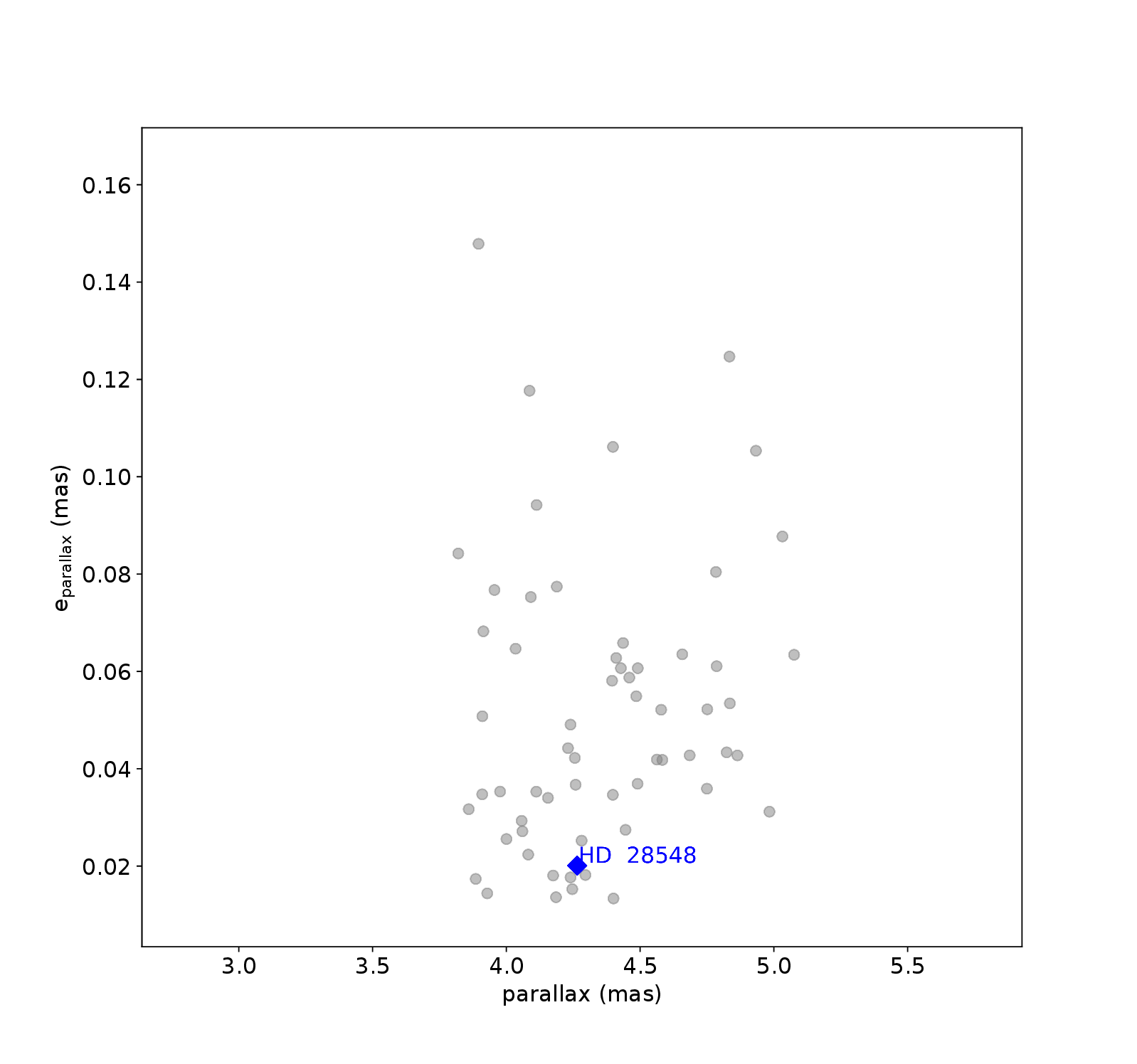}
\end{center}
\caption{Distribution of probable members of the cluster HSC 1640 (gray circles) in different spaces,
including the color$-$magnitude diagram (upper left), position (upper right), 
proper motion (lower left), and  parallax space (lower right).
The position of the $\lambda$~Boo stars is indicated in the plots (blue diamonds),
also the cluster center is marked in the coordinates space (red plus).
The best isochrone fit is shown with a red continuous line in the color$-$magnitude diagram (upper left).} 
\label{hsc1640}
\end{figure*}

\begin{figure*}
\begin{center}
\includegraphics[width=8cm]{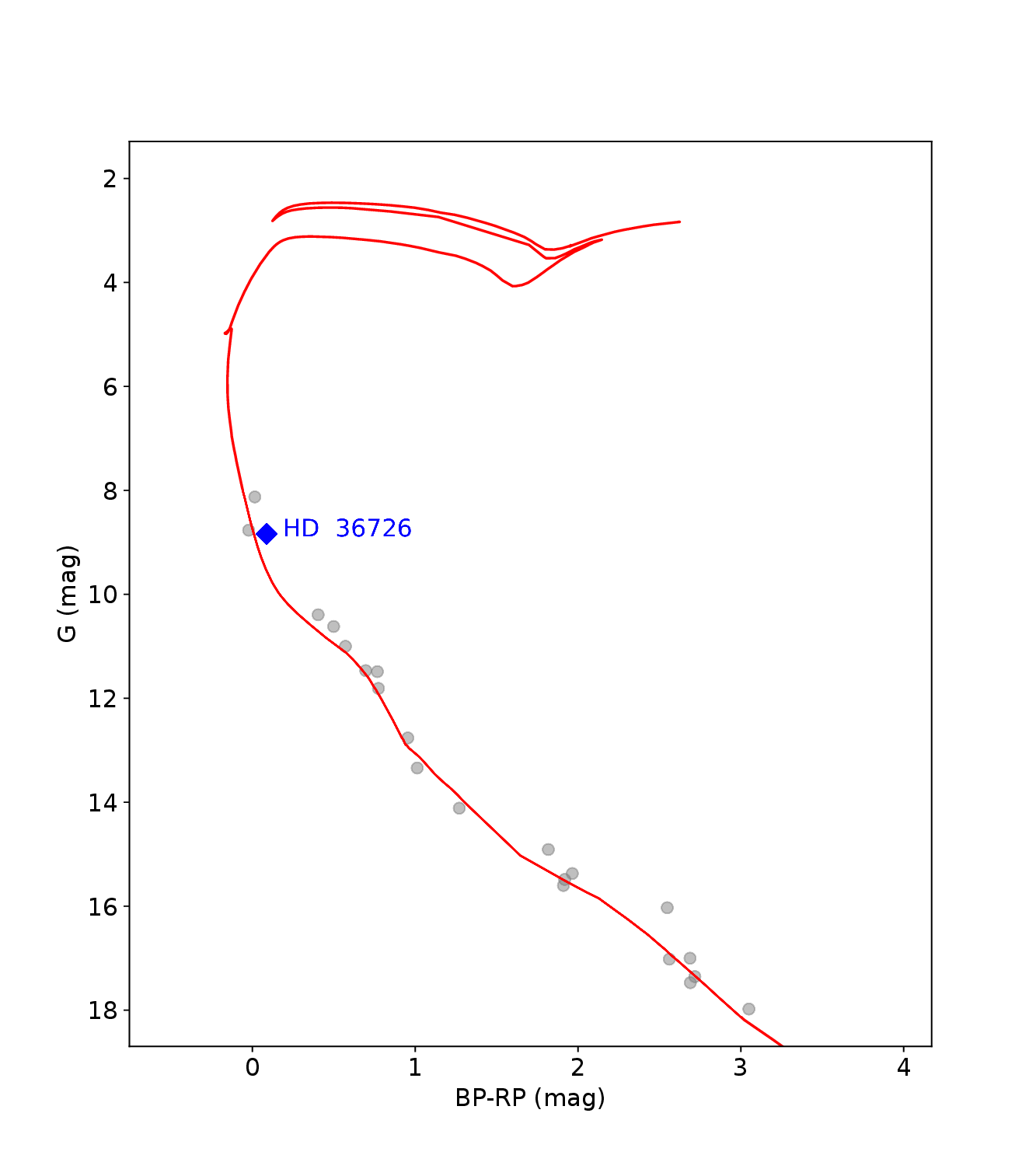}
\includegraphics[width=8cm]{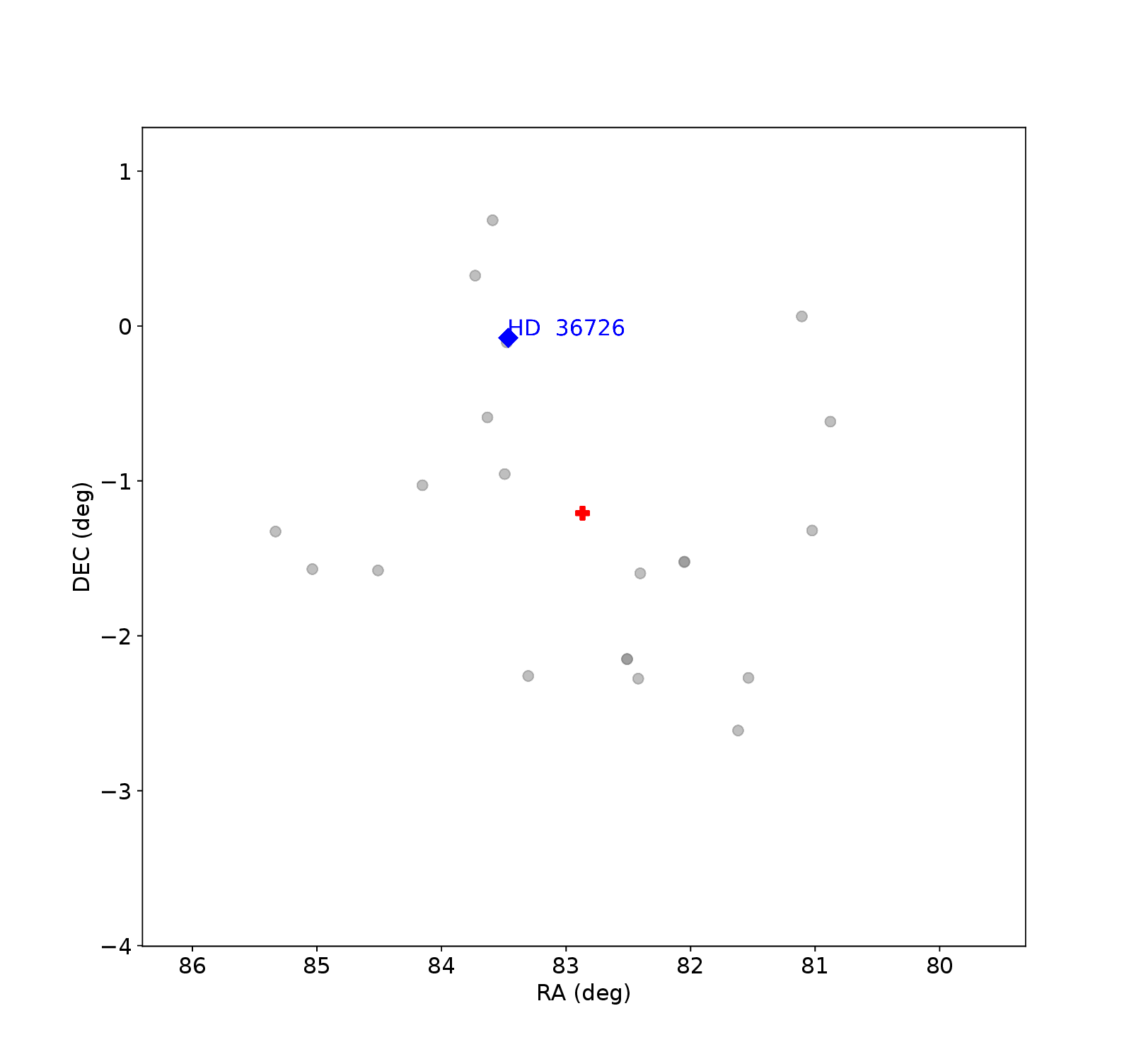}
\includegraphics[width=8cm]{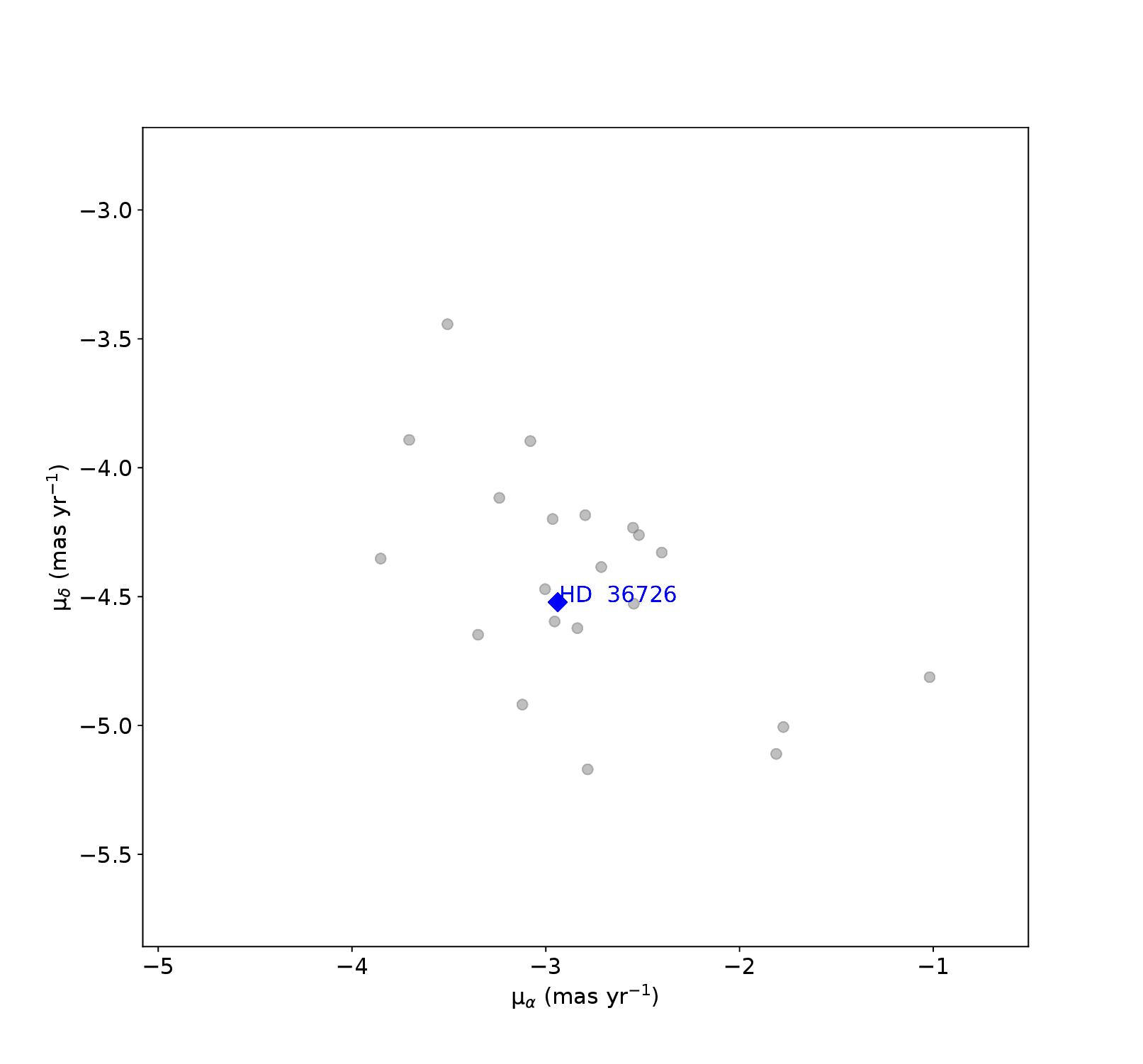} 
\includegraphics[width=8cm]{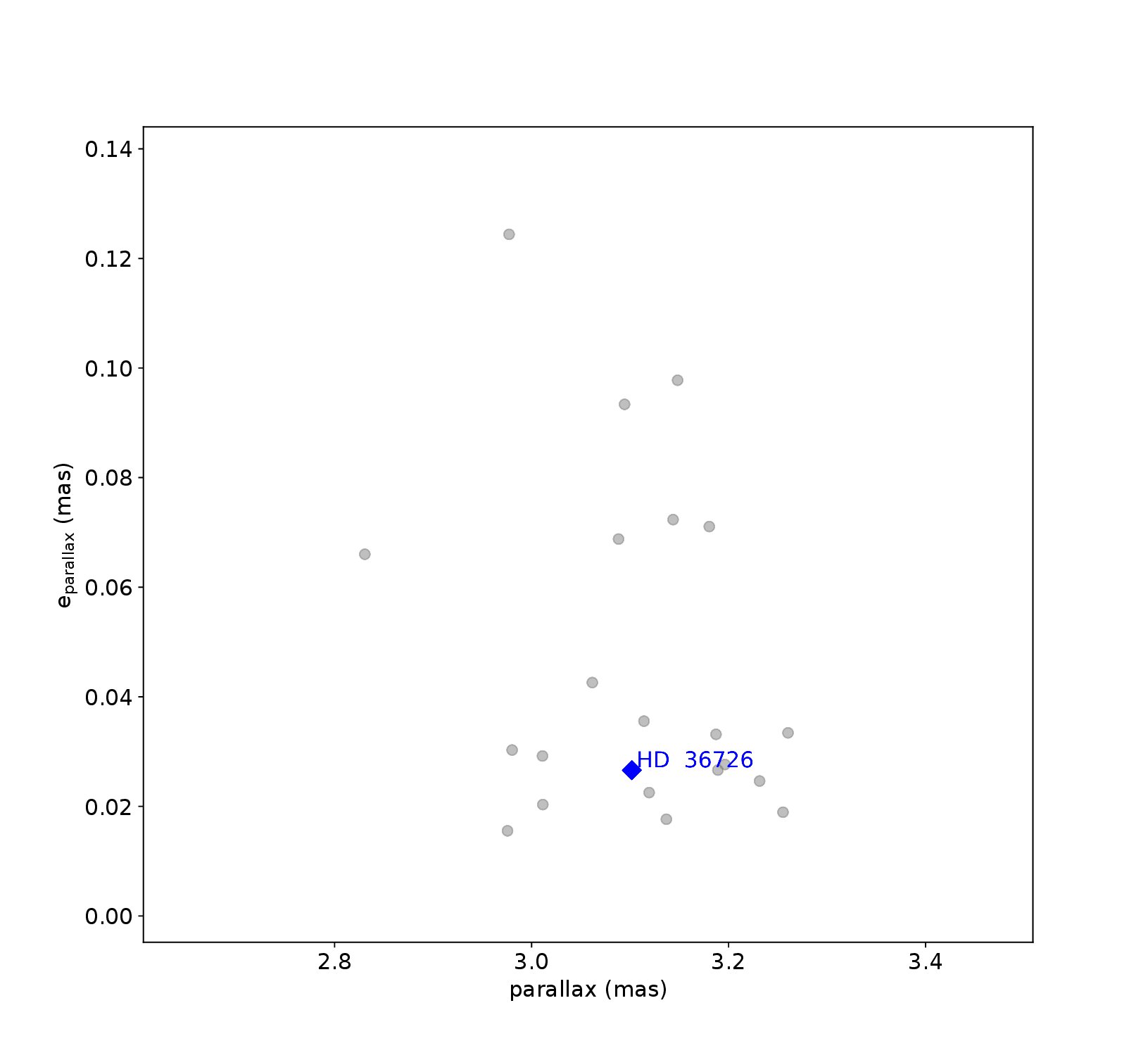}
\end{center}
\caption{Distribution of probable members of the cluster Theia 139 (gray circles) in different spaces,
including the color$-$magnitude diagram (upper left), position (upper right), 
proper motion (lower left), and  parallax space (lower right).
The position of the $\lambda$~Boo stars is indicated in the plots (blue diamonds),
also the cluster center is marked in the coordinates space (red plus).
The best isochrone fit is shown with a red continuous line in the color$-$magnitude diagram (upper left).} 
\label{theia139}
\end{figure*}

Then, we performed a photometric selection of probable members, taking into account their position in the color$-$magnitude diagram.
We made this selection and also obtained the age of the clusters using an iterative PYTHON script.
Briefly, we started the iterations by providing an approximate value of extinction and cluster age.
Then, the algorithm determines the parallax and average metallicity weighted with Gaia data.
With this information, the script selects an isochrone (that will be used to fit the photometric data)
and derive new values for the parallax and extinction.
We used a grid of PARSEC stellar models \citep{parsec, parsec2}
with isochrones between log $\tau$ $=$ 6.00 and log $\tau$ $=$ 8.85
and metallicity between [Fe/H]$=$-0.3 dex and [Fe/H]$=$+0.3 dex.
Next, the algorithm uses the bootstrap technique to randomly select and replace a number of stars
(equal to the number of probable cluster members).
In every fit, the script determines the residuals weighted by the error in magnitude and color, and selects the fit with the minimum residual.
Then, the program considers the location of the stars in the color$-$magnitude diagram to detect non-members of the cluster.
In this case, the script determines the minimum distance of the stars to the isochrone (d)
and rules out those objects with a d $>$ 0.7 mag.
The final isochrone corresponds to the one with the minimum residuals.
Figures~\ref{hsc1640} and \ref{theia139} show the $G$ vs. ($G_{BP}-G_{RP}$) color$-$magnitude diagram of probable members
of the two clusters and the isochrone fit corresponding to log $\tau$ $=$7.42$\pm$0.14 for HSC~1640 and 7.52$\pm$0.05 for Theia~139.
Finally, the program performs a new determination of parallax and extinction.

Once we had made the kinematic and photometric selection of probable members, 
we calculated the cluster parameters by averaging all stars weighted with the membership probability and its uncertainty. 
In Table~\ref{oc} we list the spectroscopic, astrometric and photometric parameters obtained for both clusters. 
Considering the observed objects, we found that
HD~28548 is a probable member of the cluster HSC~1640, while
the stars HD~36726, HD~37187 and HD 290621 are probable members of the cluster Theia~139. 
In Figures~\ref{hsc1640} and \ref{theia139}, we show the distribution of probable members (gray circles)
in different spaces, including the color$-$magnitude diagram (upper left), position (upper right), 
proper motion (lower left), and  parallax space (lower right).
The position of the $\lambda$~Boo stars is indicated in the plots (blue diamonds),
also the cluster center is marked in the coordinates space (red plus).
The best isochrone found by the script is shown with a continuous red line in the color $-$ magnitude diagram 
(upper left panel of Figs. \ref{hsc1640} and \ref{theia139}).
We can see from Figs. \ref{hsc1640} and \ref{theia139} that the $\lambda$~Boo stars are located on the periphery of the clusters.

\begin{table}
\caption{Parameters of the clusters HSC 1640 and Theia 139.} \label{oc}
\begin{center}
\begin{tabular}{ccr@{$\pm$}lr@{$\pm$}l}
\hline\hline
Parameter                        & Units  & \multicolumn{2}{c}{HSC~1640}& \multicolumn{2}{c}{Theia~139} \\ \hline
\noalign{\smallskip}
\multicolumn{6}{c}{Spectroscopic parameters} \\
\noalign{\smallskip}
RV                                 &  [km s$^{-1}$]   &    13.4      &   0.5      &    16.3        & 0.3 \\
\hline    
\noalign{\smallskip}
\multicolumn{6}{c}{Astrometric parameters} \\
\noalign{\smallskip}
$\alpha$(J2000)                  &  [deg]            &   68.18    &  0.39        & 82.87            &0.16  \\
$\delta$(J2000)                  &  [deg]            &  $-$9.61  &  0.33     & $-$1.21       &0.18  \\
$\mu_\alpha$                     &  [mas yr$^{-1}$]  &  2.09     & {0.09}  & $-$2.84       & {0.06} \\
$\mu_\delta$                     &  [mas yr$^{-1}$]  &  $-$3.73  &{0.09}  &  $-$4.54         & {0.07} \\
Parallax$_{Astr}$                &  [mas]            & { 4.25}&{0.03}     &  {3.11}       & {0.02} \\
$N_{G<18}$                       &                   &\multicolumn{2}{c}{59}&\multicolumn{2}{c}{22} \\ \hline
\noalign{\smallskip}
\multicolumn{6}{c}{Photometric parameters$^{(a)}$} \\
\noalign{\smallskip}
Parallax$_{Phot}$                &  [mas]            & { 5.01}&{0.19}     &  {2.59}       & {0.11} \\
$E(G_\mathrm{BP}-G_\mathrm{BP})$ &  [mag]            &   0.29    & 0.12      & 0.16          & 0.11    \\
log $\tau$                      &  [dex]            &    7.42    &    0.14    &  7.52         &  0.05      \\
\hline
\end{tabular}
\tablefoot{
\tablefoottext{a}{Photometric parameters derived from the isochrone fit.}
}
\end{center}
\end{table}

\section{Discussion}

\subsection{On the nature of HSC 1640 and Theia 139}

HSC 1640 is considered an open cluster in several works \citep{hunt-reffert23,cavallo24,hu24,maconi25}.
Theia 139 is also considered an open cluster in different works \citep{hunt-reffert23,cavallo24,faltova25}.
Both clusters are included in The Unified Cluster Catalog (UCC\footnote{https://ucc.ar/}).
However, \citet{hunt-reffert24} caution that both clusters are possibly dissolving.
They attempted to separate bound and unbound clusters, and identified them as "open clusters" (OCs) and "moving groups" (MGs).
First, they applied a membership algorithm with the main aim of identifying open clusters, that is, a single population,
and then they classified the clusters found in MGs and OCs.
The separation was performed by estimating the probability that a given cluster has a valid Jacobi radius.
Then, the authors consider that those clusters classified as MGs are dissolving into the disk.
In this way, they classified HSC 1640 and Theia 139 as OCs in a first work \citep{hunt-reffert23} 
and then as MGs when applying the Jacobi radius criteria in a subsequent work \citep{hunt-reffert24}.
However, they caution that the Jacobi radius method could present serious limitations in case of
low mass clusters (such as HSC 1640 and Theia 139).
For these low mass clusters, they caution that the Jacobi radii may be erroneously calculated and even show some examples.
Therefore, we consider that the classification of HSC 1640 and Theia 139 as MGs should be taken with caution.

We also note the following.
In general, MGs are usually defined as stars with similar spatial and kinematic properties \citep[e.g.][]{antoja08,antoja10}.
The origin of MGs is attributed to different mechanisms, 
such as the trapping of stars by Galactic resonances and/or the dissolution of open clusters \citep[e.g. ][]{antoja10,barros20,gagne21,yang21}.
A number of MGs display a very wide range of ages \citep[e.g.][]{antoja08,kushniruk20},
implying a mix of stellar populations.
For example, the moving group HR 1614 presents a wide range of ages and metallicity \citep{kushniruk20},
while \citet{yang21} discussed some examples of MGs which include a mixture of thin and thick disk stars.
The examples mentioned show that, in general, MGs do not represent a single population of stars.
However, this differs from the criteria adopted by \citet{hunt-reffert24} to separate MGs and OCs:
in their work, both MGs and OCs are considered a single population, separated only by the Jacobi radius criteria.
We suppose that \citet{hunt-reffert24} intended to separate bound from unbound objects,
rather than provide a precise definition of MGs.
Taking into account the mentioned limitations of the Jacobi radius criteria, 
we consider it appropriate to refer to HSC 1640 and Theia 139 as "possibly dissolving" open clusters.
Beyond the classification of HSC 1640 and Theia 139 (OCs, MGs or multiple systems),
the most relevant aspect for the study of $\lambda$ Boo stars is that both groups
belong to a single population, as showed by \citet{hunt-reffert23,hunt-reffert24}
and also verified in this work with a multi-criteria analysis (section \ref{section.membership}).

\subsection{$\lambda$ Boo stars in the cluster Theia 139}

Following our membership analysis (section \ref{section.membership} and Table \ref{vr}),
we have found three stars that belong to the cluster Theia 139:
HD 36726, HD 37187 and HD 290621.
However, the other two stars analyzed (HD 37333 and HD 290541) do not seem to belong to the same population.
In addition, our abundance analysis showed that HD 36726 exhibits a $\lambda$ Boo chemical pattern.
Then, this remarkable finding implies that the $\lambda$ Boo star HD 36726
could be analyzed together with other stellar siblings (such as HD 37187 and HD 290621),
in order to study the origin of $\lambda$ Boo stars.
We note that HD 36726 is one of the three $\lambda$ Boo stars reported by \citet{paunzen-gray97}
as members of the Orion OB1 association. On this occasion,
we confirmed its $\lambda$ Boo class and verified its membership to a stellar population.

To estimate the original chemical composition of Theia 139,
it is reasonable to consider the two non-peculiar stars in our sample, namely,
HD 37187 and HD 290621.
Both stars display mostly solar or slightly subsolar composition,
and their metallicity agree well within the errors
([FeI/H]$=$-0.09$\pm$0.15 dex and [FeI/H]$=$-0.12$\pm$0.18 dex).
The agreement between both stars supports the idea that they belong to the same population.
However, HD 37187 is significantly hotter than HD 290621
(their T$_\mathrm{eff}$ are 10500$\pm$250 K and 7000$\pm$250 K).
Considering that diffusion effects are more pronounced
in early-type stars compared to late-type objects \citep[see e.g.][]{dotter17},
HD 290621 should present, in principle, a chemical composition 
closer to that of the original cloud of Theia 139.

In order to determine the magnitude of the $\lambda$ Boo phenomena,
we can compare the $\lambda$ Boo star HD 36726 with a proxy for the
original composition of Theia 139, that is, HD 290621.
There is a significant difference in the composition of both stars,
as we can see in Figs. \ref{fig.pattern.HD36726} and \ref{fig.pattern.HD290621}.
In fact, these two stars present one of the highest differences found between two siblings
that belong to the same population.
For instance, their metallicity differ by $\sim$0.5 dex ([FeI/H]$=$-0.72$\pm$0.2 dex and [FeI/H]$=$-0.12$\pm$0.18 dex).
This strengthens the idea that $\lambda$ Boo stars were born with a near-solar composition,
similar to a recent result \citep{alacoria22,alacoria25}. 
Also, it is interesting to note that the lithium is intense in HD 290621 ([Li/H]$=$2.07$\pm$0.21 dex),
whereas it is not detected in the $\lambda$ Boo star HD 36726.
A similar observation about Li was noted by \citet{alacoria25}, who found Li in some stars but not in their $\lambda$ Boo binary companions.
We caution that the Li abundance is sensitive to different factors such as age, metallicity \citep[e.g. ][]{martos23} and
even planet engulfment events \citep[e.g. ][]{saffe17,flores24,miquelarena24}, 
which are not neccesarily related to the $\lambda$ Boo phenomena.
Regarding the abundances of other light elements,
C resulted lower in the $\lambda$ Boo star HD 36726 compared to HD 290621 ([C/H]$=$-0.80$\pm$0.18 dex and [C/H]$=$-0.13$\pm$0.18 dex),
however O resulted similar in both stars ([O/H]$=$-0.29$\pm$0.12 dex and [O/H]$=$-0.21$\pm$0.11 dex).
Thus, the $\lambda$ Boo phenomena produces a significant variation in the heavier elements,
and, perhaps, a less significant but still noticeable variation in the lighter elements,
in agreement with the result of \citet{alacoria25}.
We caution that C and O abundances were corrected by NLTE effects. 
Therefore, we cannot completely discard that NLTE effects possibly played a role in this comparison.
It would be valuable to confirm this result for the case of the cluster HSC 1640.

A number of works have explored the possibility that IR excesses could be related to $\lambda$ Boo stars.
However, there is no conclusive evidence supporting this relationship \citep[e.g.][]{paunzen03a,draper16,gray17,murphy20}.
We present in Fig. \ref{fig.seds} (right panel) an example of the observed and modeled SEDs of the star HD 36726.
Interestingly, we find an IR excess at the WISE band W4, however,
no clear excess is present in HD 37187 nor HD 290621.
Nevertheless, we caution that a larger sample of stars would help shed light on this possible relationship.

It would be interesting to determine why the $\lambda$ Boo phenomena only developed in HD 36726
but not in HD 37187 nor HD 290621. 
We showed in section \ref{section.membership} that the three stars belong to the same stellar population (Theia 139).
Let's start by exploring the T$_\mathrm{eff}$ of these stars.
For HD 36726, HD 37187 and HD 290621 we obtained T$_\mathrm{eff}$ values of 
9000$\pm$250 K, 10500$\pm$250 K and 7000$\pm$250 K.
These temperatures would approximately correspond to spectral types A2, B9 and F1
\citep[adopting the temperature scale of ][ their Table 5]{kenyon-hartmann95}.
The homogeneous list of 118 $\lambda$ Boo stars present (hydrogen) spectral types ranging
between B8/B9 and F5 \citep{murphy15,gray17,murphy20}.
Therefore, we cannot discard that the T$_\mathrm{eff}$ possibly played a role helping to avoid the 
development of the $\lambda$ Boo phenomena, specially for the hot star HD 37187.
However, the presence of 22 F1-type objects in the list of 118 $\lambda$ Boo stars,
suggests that the $\lambda$ Boo phenomena could still operate at the T$_\mathrm{eff}$ of HD 290621.
This would imply that other factor beyond the T$_\mathrm{eff}$ possibly avoided the
$\lambda$ Boo phenomena in HD 290621.
The rotational velocities vsini of HD 36726 and HD 290621 are not very different (99.6 $\pm$ 0.9 km/s and 106.4 $\pm$ 1.4 km/s).
Perhaps the lack of an IR excess (i.e. the absence of a circumstellar disk) around HD 290621 played a role.
However, the relationship between IR excess and $\lambda$ Boo stars should be taken with caution, as previously explained.

We note that the $\lambda$ Boo star HD 36726 belongs to the periphery of Theia 139
rather than to the central regions (see Fig. \ref{theia139}).
\citet{gray-corbally02} discussed the lack of $\lambda$ Boo stars in open clusters
and mentioned that few candidate $\lambda$ Boo stars are found in the 
periphery of star-forming regions (the Orion Nebula).
They suggested that this is due to the presence of a factor external to the star and
related to membership in open clusters that prevents the operation of the $\lambda$ Boo mechanism.
They later concluded that, possibly, the factor is the photoevaporation by UV radiation
from massive O-type cluster stars of the circumstellar disk material
(which otherwise would be accreted producing the $\lambda$ Boo phenomena).
In this way, they suggested that the present field $\lambda$ Boo stars would have formed
essentially in isolation either on the peripheries of these star-formation regions.
\citet{gray-corbally02} also noted that this would be in agreement with the accretion 
scenario of \citet{venn-lambert90}, a theory for the $\lambda$ Boo mechanism that requires 
the accretion of metal-depleted gas from interstellar or circumstellar material.
Simulations also show that circumstellar disks close to the cluster density centers are more strongly
affected by external photoevaporation than those further out \citep[see, e.g. Fig. 2 of ][]{huang24}.
However, we caution that the effect of external photoevaporation on circumstellar disks depends on several factors
\citep[see, e.g. the recent review of ][ and references therein]{winter-haworth22}.
For example, \citet{adams06} suggested that the effect of photoevaporation is relatively small
in small clusters with members of 100$-$1000 (such as in the case of Theia 139).
Then, although the photoevaporation scenario is encouraging (explaining the lack of
$\lambda$ Boo stars in clusters), we consider that it should be taken with caution.
\citet{gray-corbally02} claim that no other $\lambda$ Boo scenario seems able to account for
the lack of $\lambda$ Boo in open clusters and the presence of $\lambda$ Boo stars in the peripheries.

The location of early-type stars in the periphery of clusters does not seem to be the only requirement
for the development of $\lambda$ Boo stars.
We analyzed three members of Theia 139: the $\lambda$ Boo star HD 36726, as well as HD 37187 and HD 290621.
The stars HD 36726 and HD 290621 are both located in the periphery of Theia 139,
while HD 37187 is closer to the center.
However, the $\lambda$ Boo phenomena only developed in HD 36726 but not in HD 290621
(and the T$_\mathrm{eff}$ of HD 290621 does not seem to preclude the $\lambda$ Boo phenomena).
This would preliminarily suggest that peripheral location appears to be a necessary, though not sufficient, 
condition for the development of the $\lambda$ Boo peculiarity.
We have already noted that there is an IR excess in HD 36726 but not in HD 290621.
Perhaps, only HD 36726 presents a circumstellar disk which survived the photoevaporation 
by taking advantage of its peripheric location in Theia 139.
However, the possible relation between circumstellar disks and  $\lambda$ Boo stars should be taken with caution; 
it would be highly desirable to extend this analysis with a larger sample of stars belonging
to the same population.

\subsection{$\lambda$ Boo stars in the cluster HSC 1640}

As explained in the previous sections,
we found that the $\lambda$ Boo star HD 28548 is a member of the cluster HSC 1640,
confirming the membership initially suggested by \citet{hunt-reffert24}.
This is also considered a remarkable finding, providing another laboratory
(together with the cluster Theia 139) to study the origin of $\lambda$ Boo stars.
However, we also found that some stars originally suggested by the work of \citet{hunt-reffert24}
present a doubtful membership to the cluster (HD 25674 and BD-08 924), 
while other objects are considered non-members (BD-06 984 and BD-12 905, see Table \ref{vr}).
Then, it would be valuable to analyze HD 28548 alongside additional members of HSC 1640, 
similar to the analysis performed in Theia 139,
in order to study the origin of $\lambda$ Boo stars.

We note that the $\lambda$ Boo star HD 28548 belongs to the periphery rather than the central
regions of HSC 1640 (see Fig. \ref{hsc1640}),
which is similar to the case of the $\lambda$ Boo star HD 36726 in the cluster Theia 139.
Interestingly, we also note that HD 28548 presents an IR excess at WISE bands W3 and W4 (Fig. \ref{fig.seds}, left panel).
As previously mentioned, it would be valuable to study the $\lambda$ Boo star HD 28548
alongside additional members of the cluster.
This will be the topic of our next work, using spectra of other members of HSC 1640.

\subsection{$\lambda$ Boo stars in open clusters}

In this work, we obtained a precise age determination for the $\lambda$ Boo stars
HD 28548 and HD 36726, thanks to their membership to HSC 1640 and Theia 139.
We applied an iterative algorithm together with PARSEC stellar models \citep{parsec, parsec2}
and fitted the color-magnitude diagram with probable members (see Figs. \ref{hsc1640} and \ref{theia139}),
obtaining isochrone ages of log $\tau$ $=$7.42$\pm$0.14 for HSC~1640 and 
log $\tau$ $=$7.52$\pm$0.05 for Theia~139.
This corresponds to ages of 26.3$\pm$1.4 Myr and 33.1$\pm$1.1 Myr for the $\lambda$ Boo stars
HD 28548 and HD 36726, 
representing one of the more precise age determinations of $\lambda$ Boo stars.
Compared to other $\lambda$ Boo stars \citep[see, e.g., the HR diagram of ][]{murphy-paunzen17},
we consider that HD 28548 and HD 36726 belong to the group of relatively young $\lambda$ Boo stars.

We detected in this work two $\lambda$ Boo stars (HD 36726 and HD 28548) that belong to
the clusters Theia 139 and HSC 1640, and noted that both stars belong to the 
periphery rather than to the central regions of their respective clusters.
Therefore, we consider that the location of the stars within their clusters
seems to play a important role in the development of the $\lambda$ Boo peculiarity.
To determine the exact reason for this observation could be important 
to understanding the origin of $\lambda$ Boo stars.
This could help to explain the difficulty of finding $\lambda$ Boo stars in clusters
\citep[as many works have already shown, ][]{paunzen-gray97,gray-corbally98,paunzen01a,paunzen01b,gray-corbally02,paunzen03b,paunzen14}.
Also, this would support the early suggestion of \citet{gray-corbally02}, that is,
the lack of $\lambda$ Boo stars in clusters and their presence in the peripheries
is due to a factor external to the stars and related to the membership to open clusters.
Also, if the clusters are dissolving \citep[as suggested by ][]{hunt-reffert24},
this could possibly help $\lambda$ Boo stars to reach the peripheries of the clusters.
However, we also noted for the case of Theia 139 that the peripheric location is not enough
for the development of $\lambda$ Boo stars, suggesting that an additional factor is
required (because there are early-type stars in the periphery of Theia 139 that do not display
the $\lambda$ Boo phenomena).

We wonder if the mentioned additional factor could be the presence of a circumstellar disk.
The two $\lambda$ Boo stars detected in this work present an IR excess
(HD 28548 at WISE bands W3 and W4, and HD 36726 at the WISE band W4, see Fig. \ref{fig.seds}).
Thus, it seems that the development of the $\lambda$ Boo phenomena in clusters require both,
a peripheric location and an IR excess, at least for the presence of $\lambda$ Boo stars in clusters.
This would support the idea of photoevaporation in the central regions of the clusters suggested by \citet{gray-corbally02},
as well as the accretion of metal-depleted gas from interstellar or circumstellar material \citep{venn-lambert90}.
However, we mentioned that photoevaporation depends on several factors \citep[see, e.g. ][]{winter-haworth22}, while
\citet{adams06} suggested that the effect of photoevaporation is relatively small in small clusters
(such as HSC 1640 and Theia 139).
In addition, we caution that there is no clear relation between $\lambda$ Boo stars and IR excess in general \citep[e.g. ][]{paunzen03a,draper16,gray17,murphy20}.
For example, \citet{jura04} showed that the spectral shape of debris disk stars and two $\lambda$ Boo stars
differs fundamentally, using Spitzer/IRS data. The authors suggested that the SED of $\lambda$ Boo stars follows
a $\nu^{-1}$ power law, indicative of Poynting Robertson drag. However, \citet{chen09} suggested that
the debris disk around the star $\lambda$ Boo itself presents a central clearing, indicating that selective accretion of solids
onto the central star does not occur from a dusty disk.
Therefore, although the photoevaporation idea is promising, we consider that it should be taken with caution.
We do not discard that the mentioned scenarios may only work for the case of relatively young $\lambda$ Boo stars studied here.
\citet{murphy-paunzen17} suggested that there are different channels producing the $\lambda$ Boo spectra,
depending on diverse conditions such as the evolutionary state of the stars.

Another possible scenario for the origin of $\lambda$ Boo stars proposes the
interaction of early-type stars with a diffuse interstellar cloud \citep{kamp-paunzen02,mg09}.
In this scenario, underabundances are produced
by different amounts of volatile accreted material, while the more refractory species
are possibly separated and repelled from the star.
\citet{murphy-paunzen17} suggested that the lack of $\lambda$ Boo stars in intermediate-age clusters is explained
if the dominant source of accreted material is diffuse ISM clouds,
because such clouds are not observed in clusters.
For example, \citet{hunt-reffert24} cross-matched their open cluster catalog with a catalog of nearby molecular clouds \citep{cahlon24},
and found that only one cluster (HSC 598) is within 25 pc of a molecular cloud, at a separation of 8 pc.
This would support the idea that open clusters are not related to ISM clouds.
However, \citet{murphy-paunzen17} caution that the ISM clouds scenario presents some problems.
For instance, the star $\delta$ Vel is interacting with the ISM but 
it does not belong to the $\lambda$ Boo class \citep{gaspar08}.
In addition, \citet{alacoria22} analyzed the triple system HD 15165 and found
a $\lambda$ Boo star (HD 15165) accompanied by a solar composition early-type star (HD 15164), 
which is difficult to explain under the scenario of ISM clouds.

\section{Concluding remarks}

We highlight the main results of this work as follows: 

-For the first time, we present the remarkable finding of two $\lambda$ Boo stars
as members of open clusters:
HD 28548 member of HSC 1640, and HD 36726 member of Theia 139.
This was confirmed using a detailed abundance analysis,
while the cluster membership was independently analyzed using Gaia DR3 data and radial velocities.
The two open clusters constitute excellent laboratories for studying the origin of
$\lambda$ Boo stars in detail.

-We compared the $\lambda$ Boo star HD 36726 with other cluster members
and suggest that it was born with a near-solar composition.
This implies one of the highest chemical differences found between two cluster members ($\sim$0.5 dex).
In addition, we suggest that the $\lambda$ Boo peculiarity strongly depletes heavier metals,
but could slightly modify lighter abundances such as C and O,
as recently suggested by our team \citep{alacoria25}.
It would be valuable to confirm this result with additional observations.
Other stars observed in our sample present a doubtful membership to the mentioned clusters.

-We found that both $\lambda$ Boo stars belong to the periphery of their respective clusters. 
This suggests that $\lambda$ Boo stars avoid the strong photoevaporation
by UV radiation from massive stars in the central regions of the clusters, 
as previously proposed \citep{gray-corbally02}.
This would also help to explain the difficulty of finding $\lambda$ Boo stars in clusters. 
We preliminarily suggest that peripheral location appears to be a necessary, though not sufficient, 
condition for the development of the $\lambda$ Boo peculiarity (because other peripheral early-type stars
do not show the peculiarity).

-We also obtained a precise age determination for the $\lambda$ Boo stars
HD 28548 (26.3$\pm$1.4 Myr) and HD 36726 (33.1$\pm$1.1 Myr), thanks to their cluster membership.
This is one of the most precise age determinations of $\lambda$ Boo stars.
We consider that both objects belong to the group of young $\lambda$ Boo stars.

We strongly encourage to analyze additional cluster members of HSC 1640 and Theia 139.
They could provide important insights to study the origin of $\lambda$ Boo stars.

\begin{acknowledgements}
We thank the anonymous referee for constructive comments that improved the paper.
The authors thank Dr. R. Kurucz for making his codes available to us.
CS acknowledges financial support from CONICET (Argentina) through grant PIP 11220210100048CO
and the National Univ. of San Juan (Argentina) through grant CICITCA 21/E1235.
IRAF is distributed by the National Optical Astronomical Observatories, 
which is operated by the Association of Universities for Research in Astronomy, Inc., under a cooperative agreement
with the National Science Foundation.
Based on observations obtained at the international Gemini Observatory, a program of NSF NOIRLab, which is managed by
the Association of Universities for Research in Astronomy (AURA) under a cooperative agreement with the U.S. National Science Foundation on behalf
of the Gemini Observatory partnership: the U.S. National Science Foundation (United States), National Research Council (Canada), 
Agencia Nacional de Investigaci\'{o}n y Desarrollo (Chile), Ministerio de Ciencia, Tecnolog\'{i}a e Innovaci\'{o}n (Argentina), 
Minist\'{e}rio da Ci\^{e}ncia, Tec., Inova\c{c}\~{o}es e Comunica\c{c}\~{o}es (Brazil), and Korea Astronomy and Space Science Institute
(Republic of Korea).

\end{acknowledgements}

\begin{appendix}

\section{Abundances and chemical patterns}  \label{section.abundances}

In this section, we briefly review the abundances and chemical patterns
for the stars of this work.

\begin{figure}
\centering
\includegraphics[width=8.0cm]{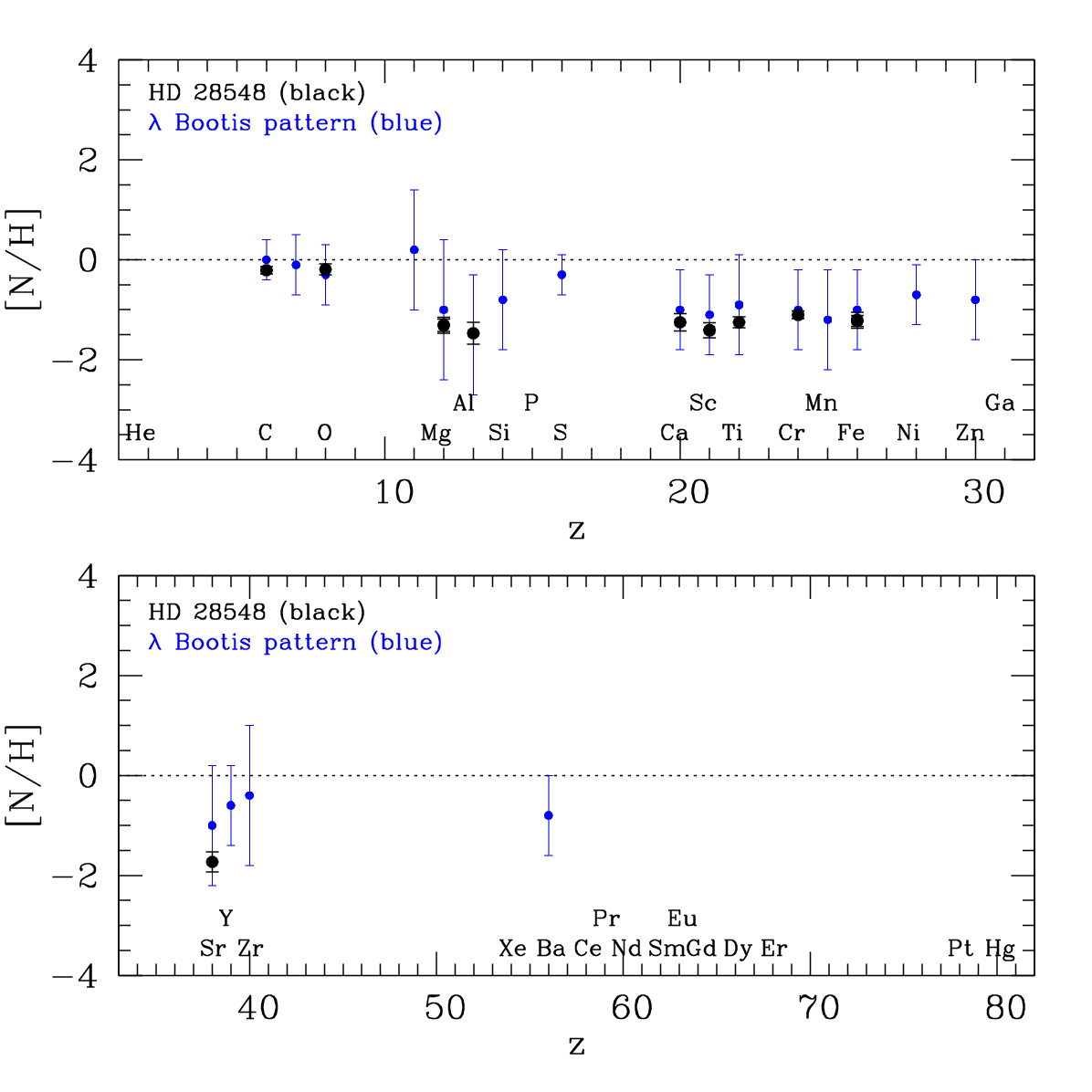}
\caption{Chemical pattern of the star HD 28548 (black), 
compared to an average pattern of $\lambda$ Boo stars (blue).}
\label{fig.pattern.HD28548}
\end{figure}

HD 28548:
This object was classified by \citet{gray17} as "A3 V kA0.5mA0.5 $\lambda$ Boo".
We present in Fig. \ref{fig.pattern.HD28548} the chemical pattern for this object,
compared to an average pattern of $\lambda$ Boo stars.
The light elements C and O exhibit nearly solar or slightly subsolar abundances
([C/H]$=$-0.21$\pm$0.07 dex, [O/H]$=$-0.19$\pm$0.11 dex).
On the other hand, heavier metals measured in this star present 
depletions of $\sim$1 dex or more compared to the Sun.
For example, [TiII/H]$=$-1.25$\pm$0.11 dex and [FeI/H]$=$-1.21$\pm$0.16 dex.
The chemical pattern shown in Fig. \ref{fig.pattern.HD28548} agrees
with those of $\lambda$ Boo stars, confirming the classification of \citet{gray17}.

\begin{figure}
\centering
\includegraphics[width=8.0cm]{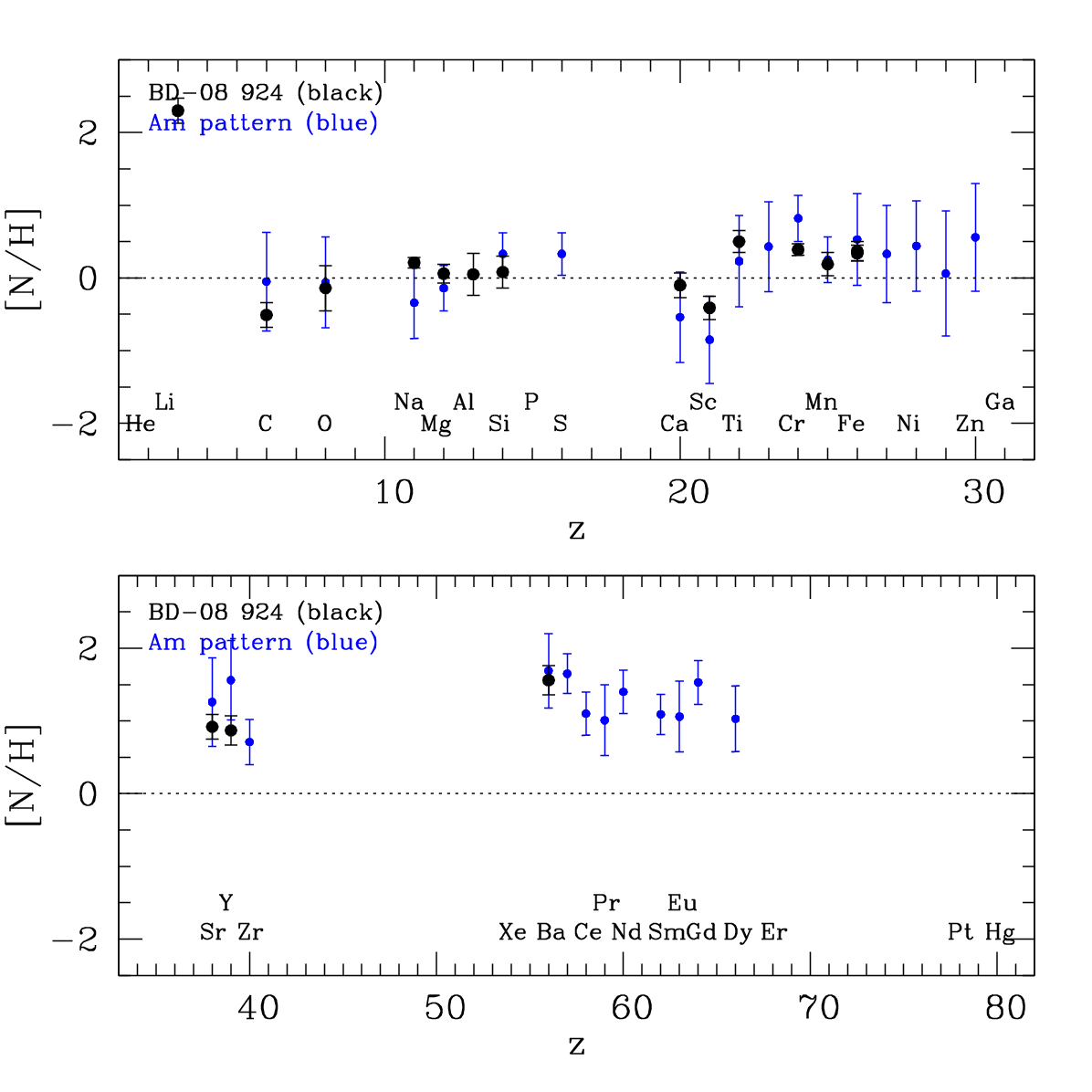}
\caption{Chemical pattern of the star BD-08 924 (black), 
compared to an average pattern of Am stars (blue).}
\label{fig.pattern.BD-08-924}
\end{figure}

BD-08 924:
Due to the metal-rich nature of this star, we present in Fig. \ref{fig.pattern.BD-08-924} its chemical pattern,
compared to an average pattern of Am stars.
This object presents subsolar values of Ca and Sc
([Ca/H]$=$-0.10$\pm$0.17 dex, [Sc/H]$=$-0.41$\pm$0.16 dex),
although slightly higher than the average of Am stars.
Other metals such as Ti, Cr, Mn and Fe exhibit suprasolar values
similar to those of Am stars.
Also, the heavier elements Sr, Y and Ba are enhanced similar to
the case of Am stars.
The Li abundance in this object is strongly above solar ($>$2 dex),
while the C abundance is relatively low 
(-0.51$\pm$0.17 dex, however, we caution that it was derived from one line,
possibly blended).
In our view, its general pattern is in agreement with that observed in Am stars.

\begin{figure}
\centering
\includegraphics[width=8.0cm]{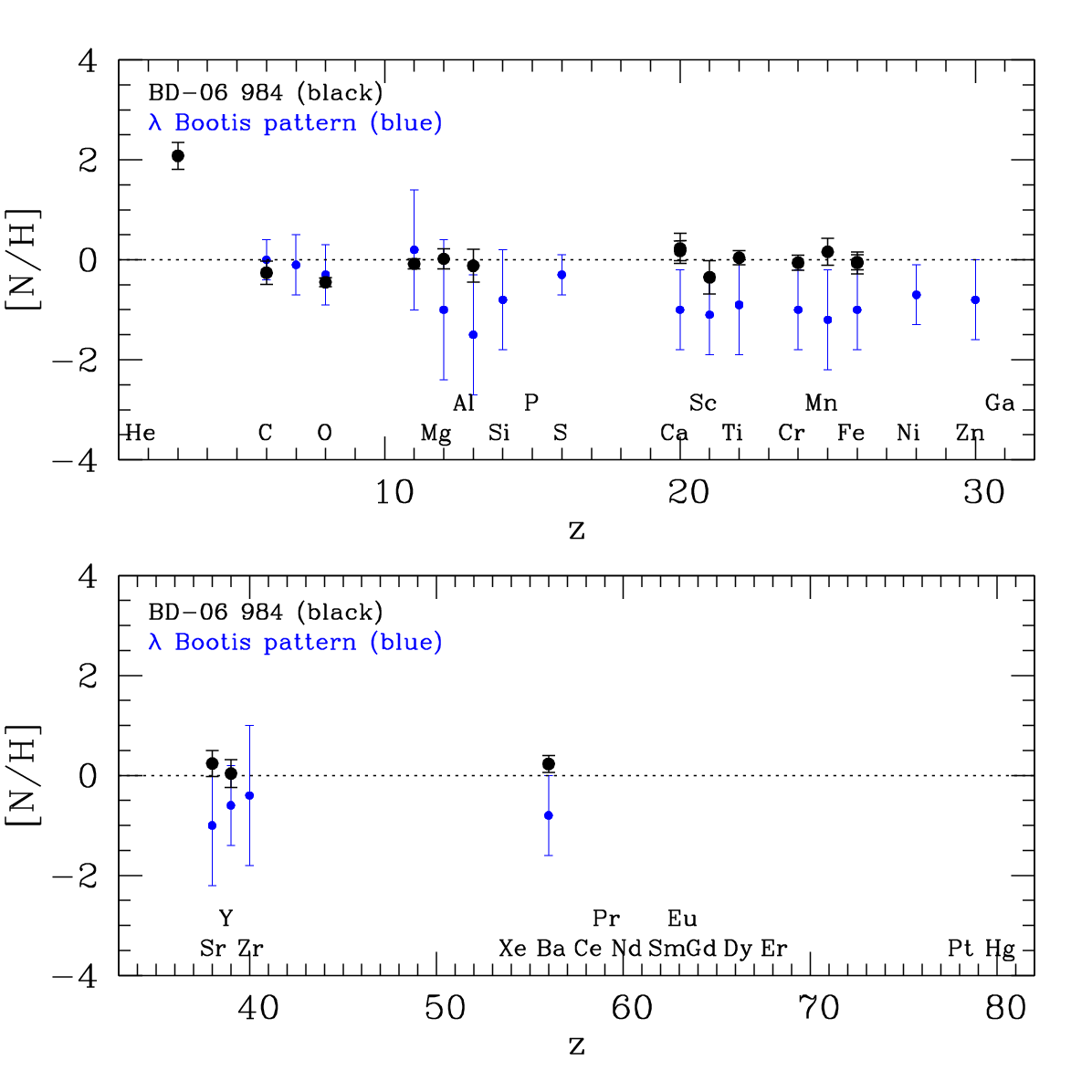}
\caption{Chemical pattern of the star BD-06 984 (black), 
compared to an average pattern of $\lambda$ Boo stars (blue).}
\label{fig.pattern.BD-06-984}
\end{figure}

BD-06 984:
We present in Fig. \ref{fig.pattern.BD-06-984} the chemical pattern for this object,
compared to an average pattern of $\lambda$ Boo stars.
Most elements exhibit nearly solar abundance values
(for instance, [FeI/H]$=$-0.06$\pm$0.22 dex).
We note that Li is strongly enhanced compared to the Sun ([LiI/H]$=$2.08$\pm$0.27 dex),
while C and O are slightly subsolar
[C/H]$=$-0.26$\pm$0.23 dex, [O/H]$=$-0.45$\pm$0.09 dex).
The general pattern presents nearly solar values, except for a few species such as Li.

\begin{figure}
\centering
\includegraphics[width=8.0cm]{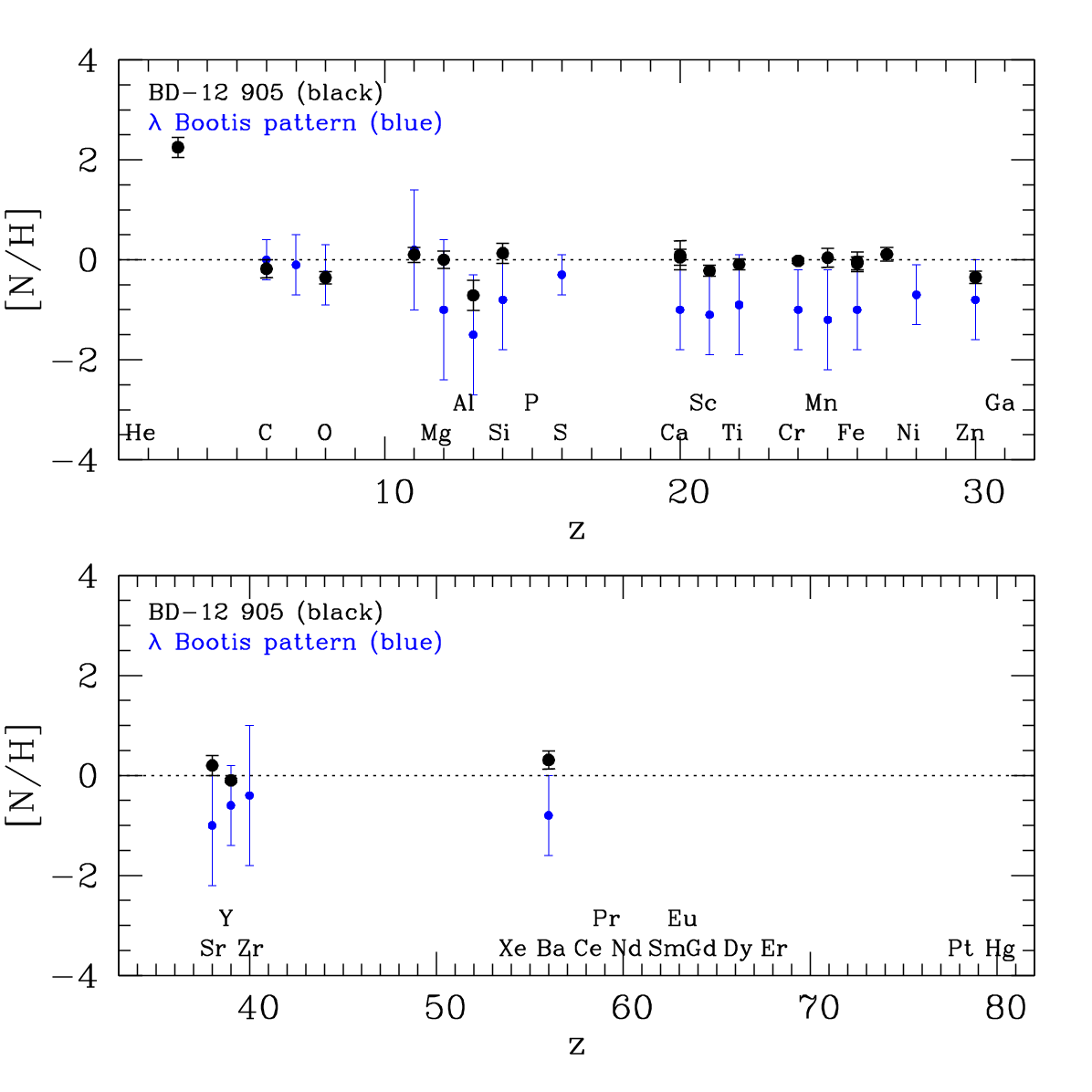}
\caption{Chemical pattern of the star BD-12 905 (black), 
compared to an average pattern of $\lambda$ Boo stars (blue).}
\label{fig.pattern.BD-12-905}
\end{figure}

BD-12 905:
We present in Fig. \ref{fig.pattern.BD-12-905} the chemical pattern for this object,
compared to an average pattern of $\lambda$ Boo stars.
Most elements exhibit nearly solar abundance values
(for instance, [FeI/H]$=$-0.04$\pm$0.20 dex).
We note that Li is strongly enhanced compared to the Sun ([LiI/H]$=$2.25$\pm$0.20 dex),
while C and O are nearly solar or slightly subsolar
[C/H]$=$-0.18$\pm$0.18 dex, [O/H]$=$-0.36$\pm$0.12 dex).
The general pattern presents mostly solar values, except for a few species such as Li.

\begin{figure}
\centering
\includegraphics[width=8.0cm]{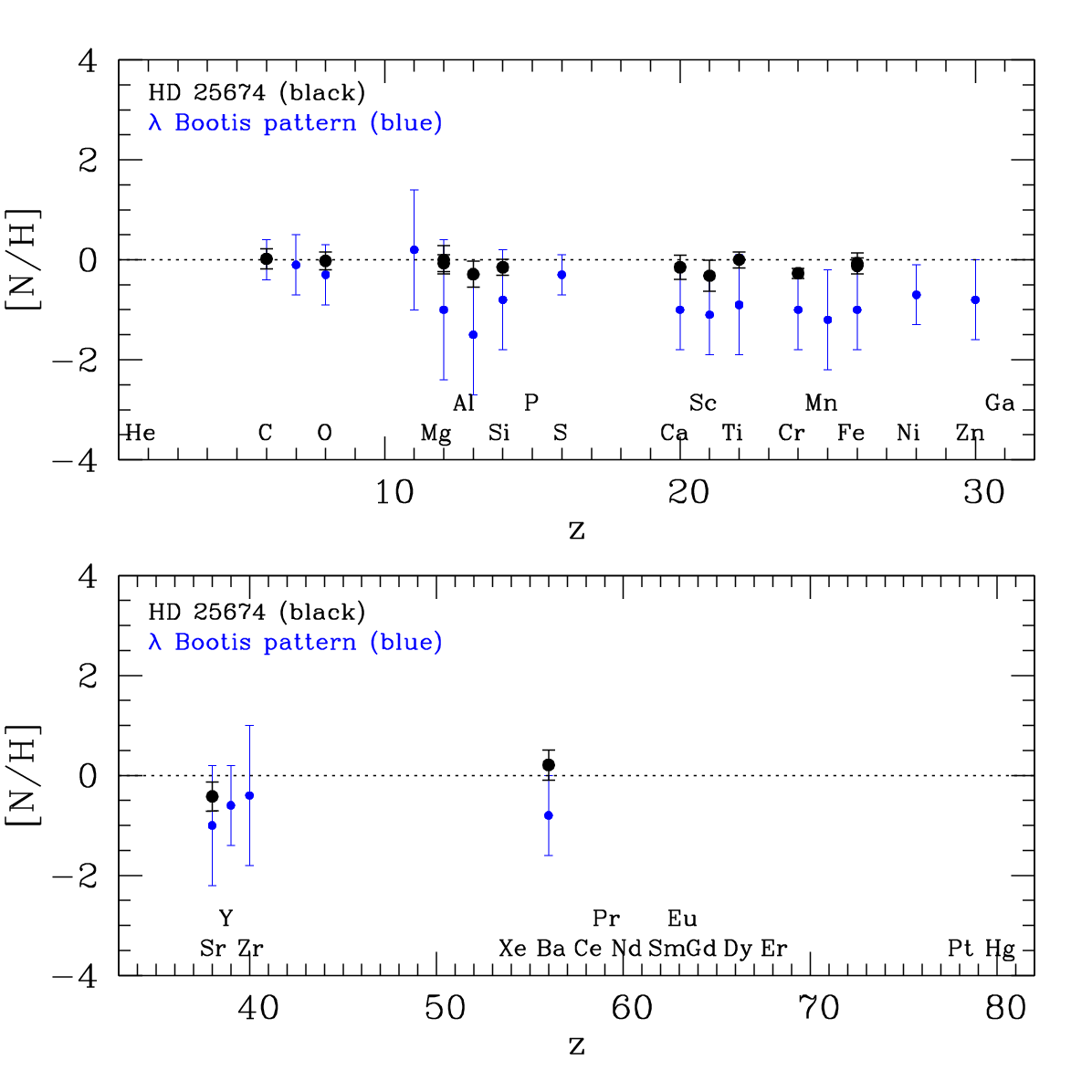}
\caption{Chemical pattern of the star HD 25674 (black), 
compared to an average pattern of $\lambda$ Boo stars (blue).}
\label{fig.pattern.HD25674}
\end{figure}

HD 25674:
We present in the Fig. \ref{fig.pattern.HD25674} the chemical pattern for this object,
compared to an average pattern of $\lambda$ Boo stars.
Most elements exhibit nearly solar abundance values
(for instance, [FeI/H]$=$-0.07$\pm$0.21 dex).
The Li line near 6707 \AA\ was not detected in the spectra of this object.
Other light elements such as C and O show solar abundance values.
The general pattern presents mostly solar values.

\begin{figure}
\centering
\includegraphics[width=8.0cm]{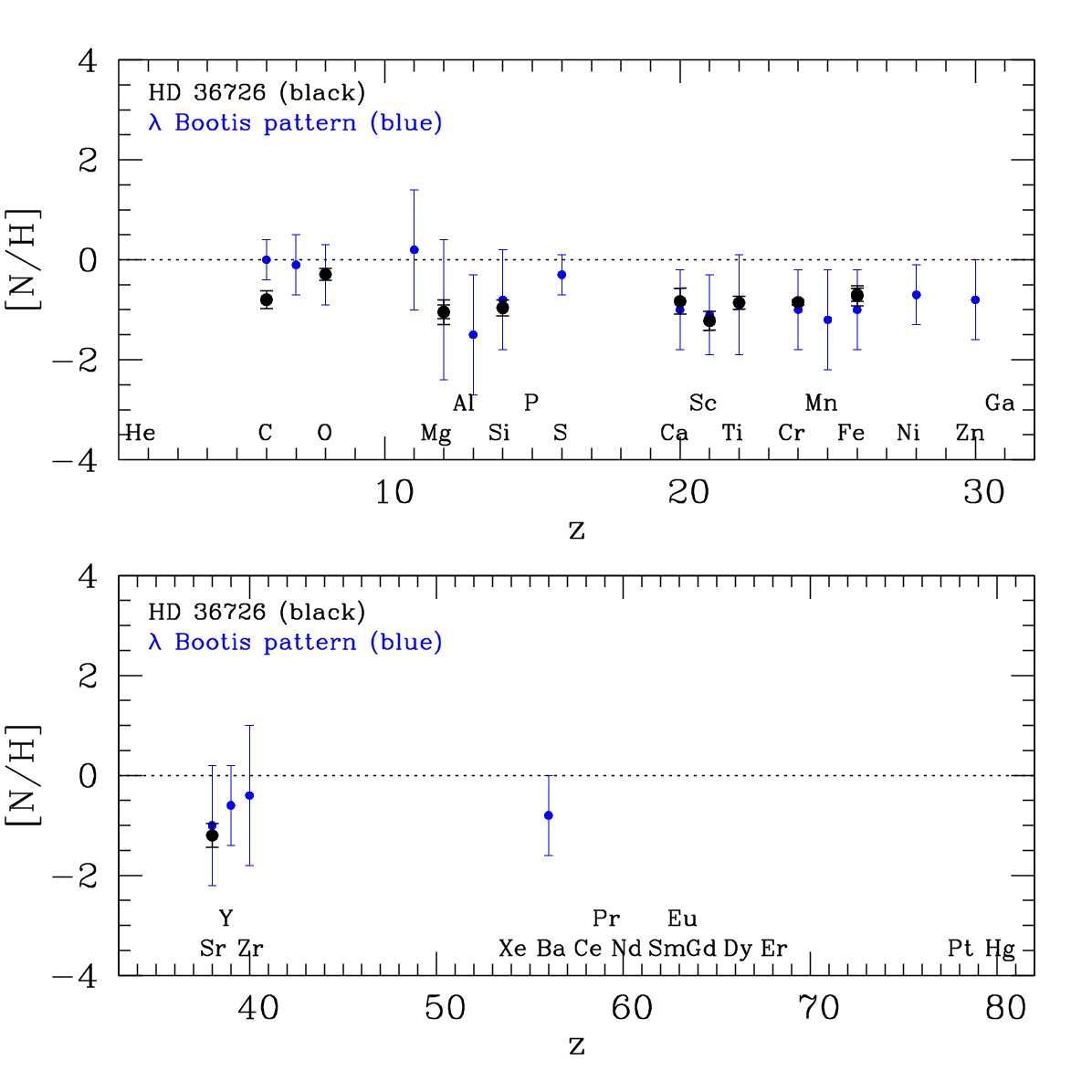}
\caption{Chemical pattern of the star HD 36726 (black), 
compared to an average pattern of $\lambda$ Boo stars (blue).}
\label{fig.pattern.HD36726}
\end{figure}

HD 36726:
This star was classified as "kA0hA5mA0 V $\lambda$ Boo" by \citet{paunzen-gray97}
and then as "hA4 Vb kA0.5mA0.5 $\lambda$ Boo" \citep{murphy15}.
We present in Fig. \ref{fig.pattern.HD36726} the chemical pattern for this object,
compared to an average pattern of $\lambda$ Boo stars.
The C abundance ([CI/H]$=$-0.80$\pm$0.18 dex) is somewhat lower than those of 
average $\lambda$ Boo stars.
Carbon lines such as 4771.74 \AA\ and 5380.34 \AA\ are relatively weak in the spectra;
then, we derived the abundance only from the clearest line at 5052.17 \AA. 
We note that the NLTE correction (NLTE-LTE) for CI amounts to -0.13 dex, 
which also diminish the CI abundance.
The O abundance is nearly solar ([OI/H]$=$-0.29$\pm$0.12 dex),
very similar to those found in $\lambda$ Boo stars.
Other metals such as Ti, Cr and Fe are depleted by $\sim$0.85 dex compared to the Sun,
which closely follow those of $\lambda$ Boo stars.
Although this chemical pattern is less extreme than those of HD 28548,
we consider it consistent with that of $\lambda$ Boo stars,
confirming the previous classification of this object \citep{paunzen-gray97,murphy15}.

\begin{figure}
\centering
\includegraphics[width=8.0cm]{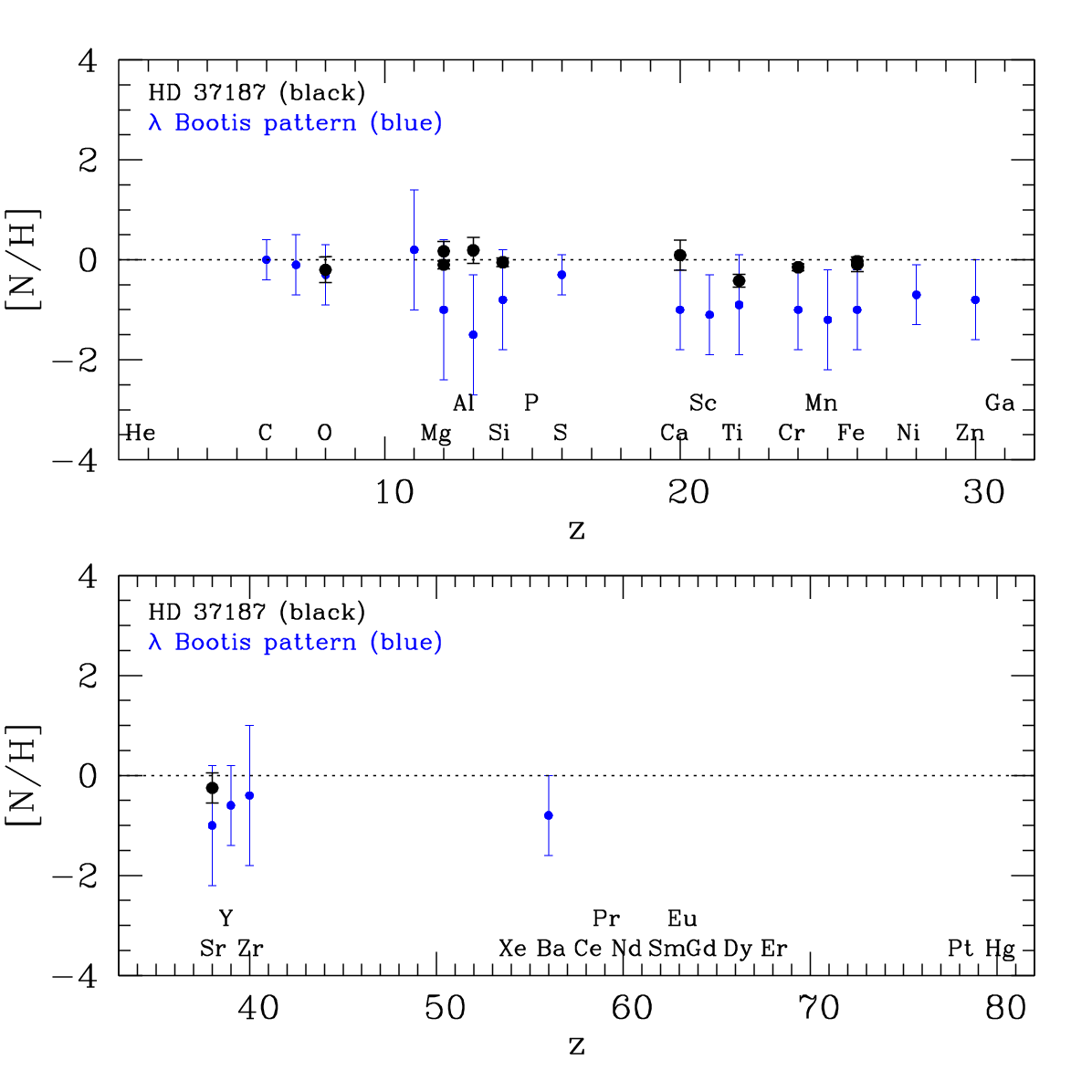}
\caption{Chemical pattern of the star HD 37187 (black), 
compared to an average pattern of $\lambda$ Boo stars (blue).}
\label{fig.pattern.HD37187}
\end{figure}

HD 37187:
We present in Fig. \ref{fig.pattern.HD37187} the chemical pattern for this object,
compared to an average pattern of $\lambda$ Boo stars.
Most elements exhibit nearly solar abundance values
(for instance, [FeI/H]$=$-0.09$\pm$0.15 dex).
The Li line near 6707 \AA\ was not detected in the spectra of this object,
while O presents a solar or slightly subsolar value.
The general pattern presents mostly solar values.

\begin{figure}
\centering
\includegraphics[width=8.0cm]{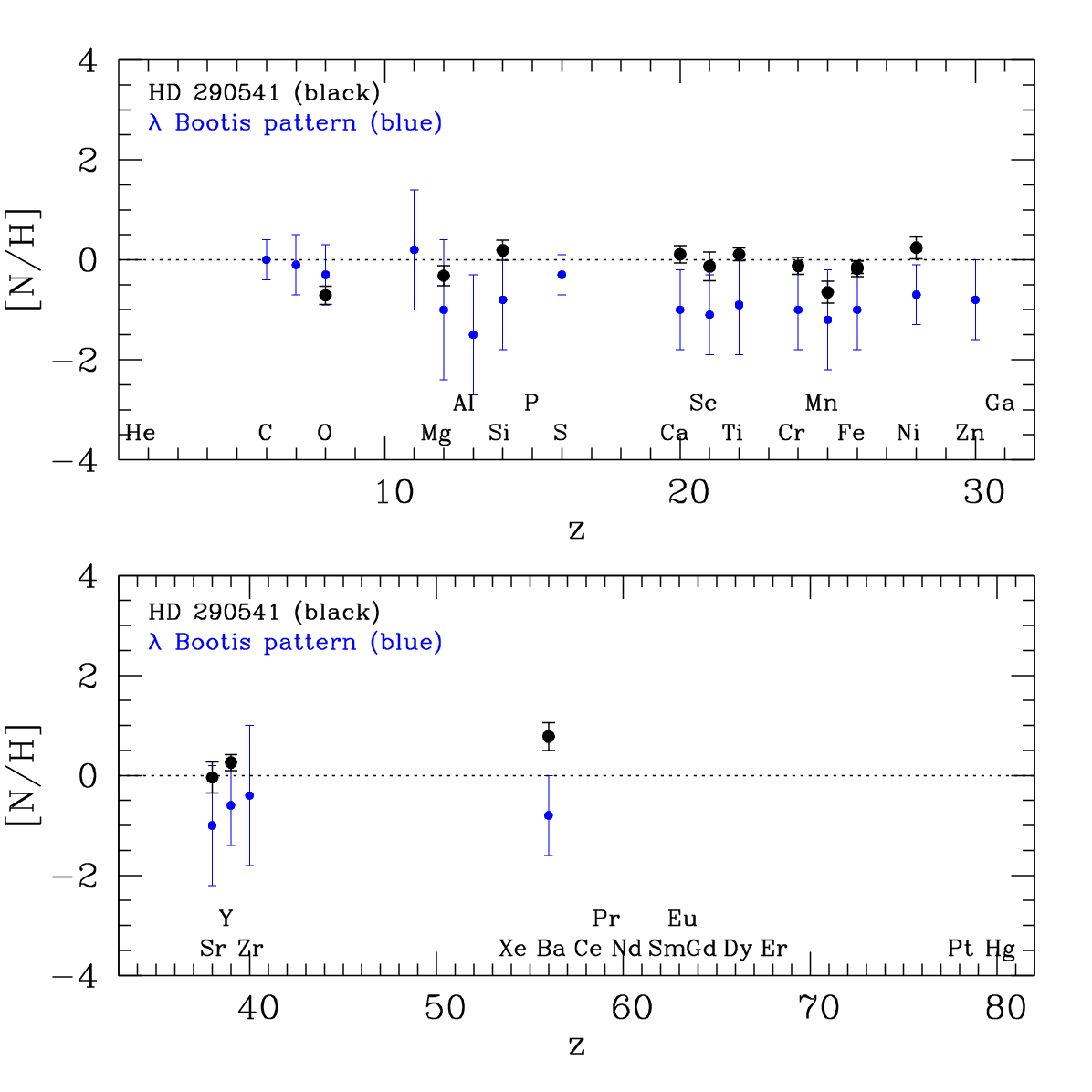}
\caption{Chemical pattern of the star HD 290541 (black), 
compared to an average pattern of $\lambda$ Boo stars (blue).}
\label{fig.pattern.HD290541}
\end{figure}

HD 290541:
We present in Fig. \ref{fig.pattern.HD290541} the chemical pattern for this object,
compared to an average pattern of $\lambda$ Boo stars.
Most elements exhibit nearly solar abundance values,
with some of them showing slightly subsolar values (O, Mg, Mn, Fe),
and other elements showing slightly suprasolar values (Ca, Ti, Ni, Ba).
The Li line near 6707 \AA\ was not detected in the spectra of this object.
We consider that the general pattern of this star presents mostly solar values
within $\pm$0.25 dex, except for a few species such as O and Ba.

\begin{figure}
\centering
\includegraphics[width=8.0cm]{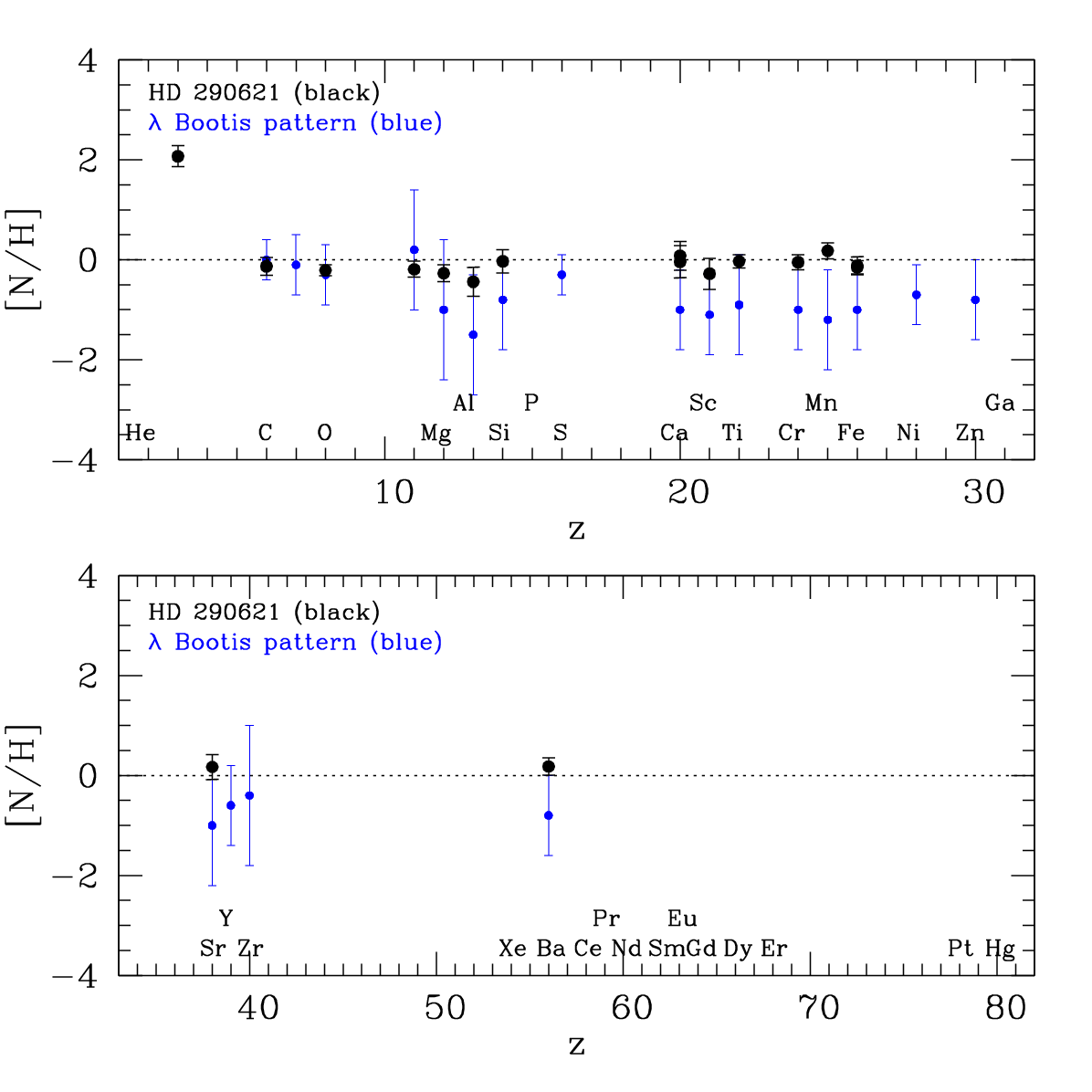}
\caption{Chemical pattern of the star HD 290621 (black), 
compared to an average pattern of $\lambda$ Boo stars (blue).}
\label{fig.pattern.HD290621}
\end{figure}

HD 290621:
We present in Fig. \ref{fig.pattern.HD290621} the chemical pattern for this object,
compared to an average pattern of $\lambda$ Boo stars.
Most elements show nearly solar abundance values
(for instance, [FeI/H]$=$-0.12$\pm$0.18 dex).
We note that Li is strongly enhanced compared to the Sun ([LiI/H]$=$2.07$\pm$0.21 dex),
while C and O are nearly solar or slightly subsolar
[C/H]$=$-0.13$\pm$0.18 dex, [O/H]$=$-0.21$\pm$0.11 dex).
The general pattern presents mostly solar values, except for a few species such as Li.

\begin{figure*}
\centering
\includegraphics[width=8.0cm]{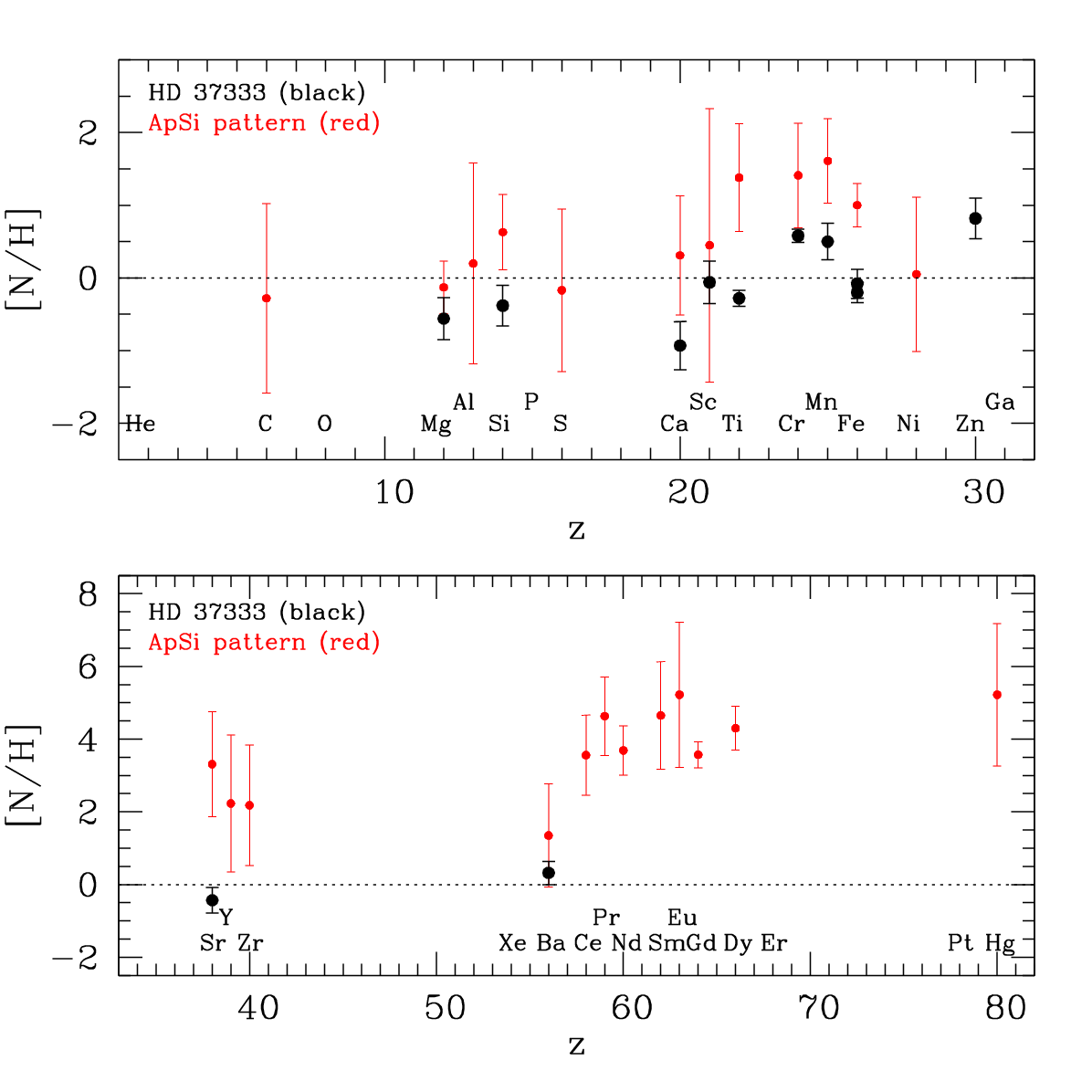}
\includegraphics[width=8.0cm]{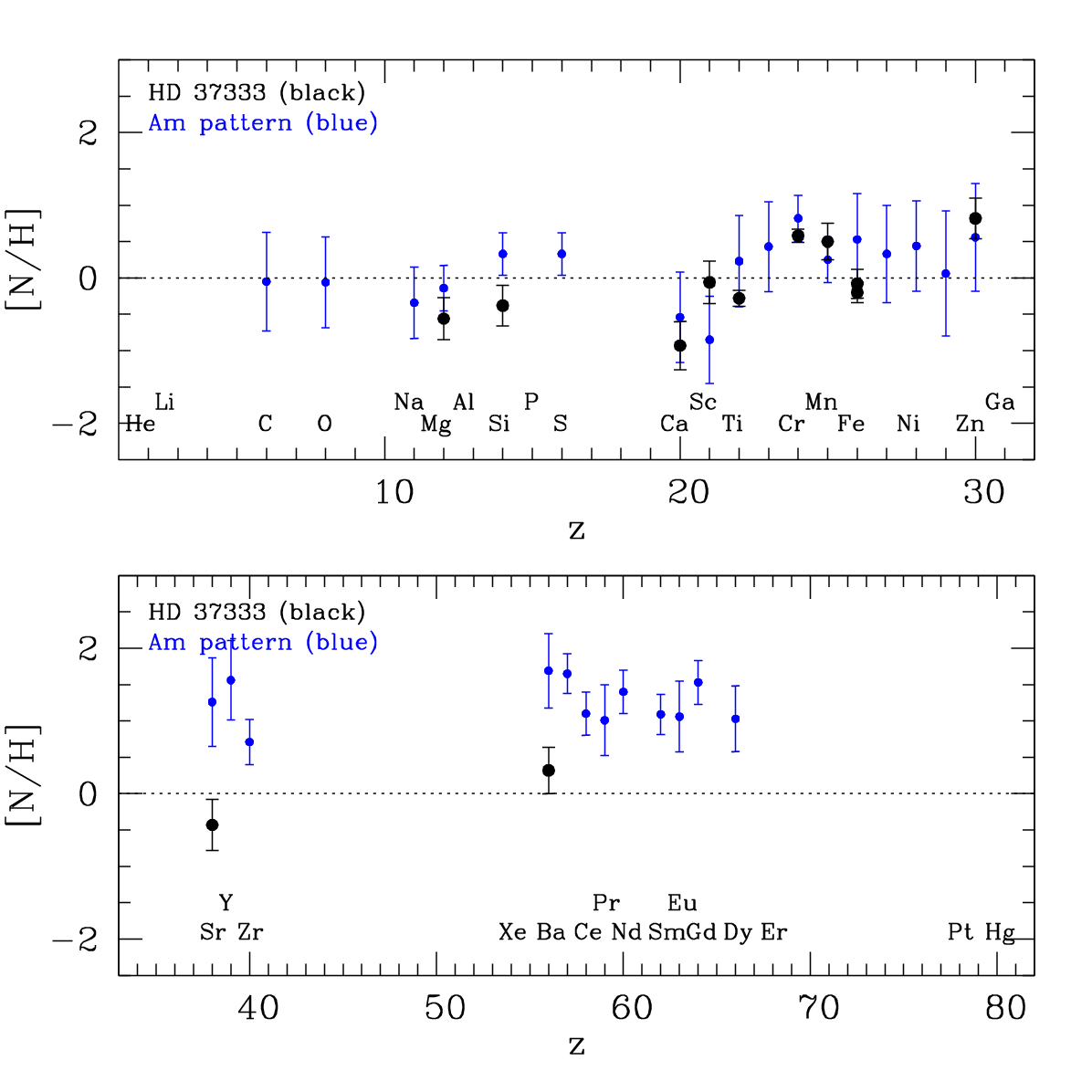}
\caption{Chemical pattern of the star HD 37333 (black), 
compared to an average pattern of Ap stars (red, left panels) and 
to an average pattern of Am stars (blue, right panels).}
\label{fig.pattern.HD37333}
\end{figure*}

HD 37333:
\citet{gray-corbally93} obtained a 2.8 \AA\ resolution spectra for this object
and classified as "A1 Va (Si II)", noting a mild-enhancement of Si II $\lambda\lambda$ 4128-30.
They consider this object as a "mild Ap star".
We present in Fig. \ref{fig.pattern.HD37333} (left panels) the chemical pattern
for this object, compared to an average pattern of ApSi stars.
We note that the Si abundance resulted subsolar ([SiII/H]$=$-0.38$\pm$0.28 dex),
different than ApSi stars.
Also, species such as Ca, Sc, Ti and Fe resulted subsolar or nearly solar, 
while Ap stars show significant enhancements of these species.
Then, we decided to compare its chemical pattern to an average pattern of Am stars
in Fig. \ref{fig.pattern.HD37333} (right panels).
Calcium is notably subsolar ([CaII/H]$=$-0.93$\pm$0.33 dex), however Sc is almost solar
([ScII/H]$=$-0.06$\pm$0.29 dex), which is different than most Am stars.
We note that Cr, Mn, Zn are enhanced compared to the Sun, similar to Am stars,
however Ti and Fe are solar or slightly subsolar.
Then, some elements seem to agree with those of Am stars, but not others (including Fe).
The T$_\mathrm{eff}$ of this object could correspond to an Am or Ap star.
Similarly, its relatively low rotational velocity (vsini $\sim$ 43.6 km/s)
do not exclude the possibility of a chemically peculiar star.
In our opinion, the pattern shown in Fig. \ref{fig.pattern.HD37333} is closer
to an Am star rather than to an Ap star.
This puzzling chemical pattern corresponds, perhaps, to a nascent or developing Am star.

\FloatBarrier

\FloatBarrier

\section{Tables of chemical abundances}   \label{section.tables}

In this section, we present the chemical abundances derived in this work and their errors.
As previously explained, 
the total error e$_{tot}$ was derived as the quadratic sum of the line-to-line dispersion e$_{1}$
(estimated as $\sigma/\sqrt{n}$ , where $\sigma$ is the standard deviation)
and the error in the abundances (e$_{2}$, e$_{3}$, and e$_{4}$) when varying T$_{\rm eff}$, $\log g$,
and v$_\mathrm{micro}$ by their corresponding uncertainties\footnote{We adopt a minimum of 0.01 dex for the errors e$_{2}$, e$_{3}$, and e$_{4}$.}.
For chemical species with only one line, we adopted as $\sigma$ the standard deviation of iron lines.
Abundance tables show the average abundance and the total error e$_{tot}$, together with
the errors e$_{1}$ to e$_{4}$.

\FloatBarrier

\begin{table}
\centering
\caption{Chemical abundances for HD 28548.}
%\hskip -0.10in
\scriptsize
\begin{tabular}{lrcccc}
\hline
\hline
Specie     & [X/H] $\pm$ e$_{tot}$ & e$_{1}$ & e$_{2}$ & e$_{3}$ & e$_{4}$ \\
\hline
C I     & -0.21 $\pm$ 0.07 & 0.06 & 0.04 & 0.01 & 0.01 \\ 
O I     & -0.19 $\pm$ 0.11 & 0.02 & 0.04 & 0.01 & 0.1 \\ 
Mg I    & -1.31 $\pm$ 0.16 & 0.04 & 0.15 & 0.01 & 0.06 \\ 
Mg II   & -1.31 $\pm$ 0.12 & 0.1 & 0.05 & 0.02 & 0.03 \\ 
Al I    & -1.47 $\pm$ 0.22 & 0.1 & 0.19 & 0.01 & 0.03 \\ 
Ca II   & -1.25 $\pm$ 0.17 & 0.1 & 0.13 & 0.01 & 0.01 \\ 
Sc II   & -1.41 $\pm$ 0.15 & 0.1 & 0.11 & 0.01 & 0.02 \\ 
Ti II   & -1.25 $\pm$ 0.11 & 0.03 & 0.1 & 0.01 & 0.03 \\ 
Cr II   & -1.1 $\pm$ 0.08 & 0.07 & 0.02 & 0.03 & 0.01 \\ 
Fe I    & -1.21 $\pm$ 0.16 & 0.02 & 0.16 & 0.01 & 0.04 \\ 
Fe II   & -1.22 $\pm$ 0.11 & 0.03 & 0.1 & 0.01 & 0.02 \\ 
Sr II   & -1.73 $\pm$ 0.2 & 0.1 & 0.17 & 0.01 & 0.02 \\ 
\hline
\end{tabular}
\normalsize
\label{tab.abunds.HD28548}
\end{table}

\begin{table}
\centering
\caption{Chemical abundances for HD 25674.}
%\hskip -0.10in
\scriptsize
\begin{tabular}{lrcccc}
\hline
\hline
Specie     & [X/H] $\pm$ e$_{tot}$ & e$_{1}$ & e$_{2}$ & e$_{3}$ & e$_{4}$ \\
\hline
C I     & 0.02 $\pm$ 0.2 & 0.15 & 0.12 & 0.01 & 0.01 \\ 
O I     & -0.02 $\pm$ 0.18 & 0.15 & 0.06 & 0.01 & 0.08 \\ 
Mg I    & 0 $\pm$ 0.28 & 0.21 & 0.18 & 0.03 & 0.05 \\ 
Mg II   & -0.07 $\pm$ 0.17 & 0.15 & 0.03 & 0.01 & 0.06 \\ 
Al I    & -0.29 $\pm$ 0.26 & 0.15 & 0.18 & 0.01 & 0.1 \\ 
Si II   & -0.15 $\pm$ 0.16 & 0.06 & 0.14 & 0.02 & 0.04 \\ 
Ca II   & -0.15 $\pm$ 0.24 & 0.15 & 0.19 & 0.01 & 0.01 \\ 
Sc II   & -0.32 $\pm$ 0.31 & 0.15 & 0.26 & 0.03 & 0.04 \\ 
Ti II   & 0 $\pm$ 0.16 & 0.05 & 0.11 & 0.01 & 0.1 \\ 
Cr II   & -0.27 $\pm$ 0.1 & 0.02 & 0.1 & 0.01 & 0.02 \\ 
Fe I    & -0.07 $\pm$ 0.21 & 0.03 & 0.19 & 0.01 & 0.06 \\ 
Fe II   & -0.12 $\pm$ 0.16 & 0.04 & 0.13 & 0.01 & 0.08 \\ 
Sr II   & -0.42 $\pm$ 0.29 & 0.16 & 0.19 & 0.02 & 0.15 \\ 
Ba II   & 0.21 $\pm$ 0.3 & 0.18 & 0.24 & 0.01 & 0.02 \\ 
\hline
\end{tabular}
\normalsize
\label{tab.abunds.HD25674}
\end{table}

\begin{table}
\centering
\caption{Chemical abundances for BD-06 984.}
%\hskip -0.10in
\scriptsize
\begin{tabular}{lrcccc}
\hline
\hline
Specie     & [X/H] $\pm$ e$_{tot}$ & e$_{1}$ & e$_{2}$ & e$_{3}$ & e$_{4}$ \\
\hline
Li I    & 2.08 $\pm$ 0.27 & 0.23 & 0.08 & 0.08 & 0.08 \\ 
C I     & -0.26 $\pm$ 0.23 & 0.23 & 0.02 & 0.01 & 0.01 \\ 
O I     & -0.45 $\pm$ 0.09 & 0.01 & 0.07 & 0.03 & 0.04 \\ 
Na I    & -0.08 $\pm$ 0.1 & 0.03 & 0.09 & 0.01 & 0.01 \\ 
Mg I    & 0.02 $\pm$ 0.2 & 0.16 & 0.13 & 0.02 & 0.03 \\ 
Al I    & -0.12 $\pm$ 0.33 & 0.23 & 0.23 & 0.02 & 0.04 \\ 
Ca I    & 0.23 $\pm$ 0.3 & 0.23 & 0.19 & 0.02 & 0.05 \\ 
Ca II   & 0.18 $\pm$ 0.2 & 0.08 & 0.18 & 0.01 & 0.01 \\ 
Sc II   & -0.35 $\pm$ 0.33 & 0.23 & 0.11 & 0.01 & 0.21 \\ 
Ti II   & 0.04 $\pm$ 0.14 & 0.04 & 0.1 & 0.01 & 0.08 \\ 
Cr II   & -0.06 $\pm$ 0.15 & 0.14 & 0.01 & 0.01 & 0.05 \\ 
Mn I    & 0.16 $\pm$ 0.27 & 0.25 & 0.07 & 0.01 & 0.06 \\ 
Fe I    & -0.06 $\pm$ 0.22 & 0.05 & 0.2 & 0.01 & 0.09 \\ 
Fe II   & -0.05 $\pm$ 0.15 & 0.05 & 0.12 & 0.01 & 0.08 \\ 
Sr II   & 0.24 $\pm$ 0.26 & 0.05 & 0.24 & 0.01 & 0.08 \\ 
Y II    & 0.04 $\pm$ 0.28 & 0.23 & 0.16 & 0.02 & 0.03 \\ 
Ba II   & 0.23 $\pm$ 0.17 & 0.09 & 0.12 & 0.01 & 0.08 \\ 
\hline
\end{tabular}
\normalsize
\label{tab.abunds.BD-06-984}
\end{table}

\begin{table}
\centering
\caption{Chemical abundances for BD-08 924.}
%\hskip -0.10in
\scriptsize
\begin{tabular}{lrcccc}
\hline
\hline
Specie     & [X/H] $\pm$ e$_{tot}$ & e$_{1}$ & e$_{2}$ & e$_{3}$ & e$_{4}$ \\
\hline
Li I    & 2.3 $\pm$ 0.17 & 0.16 & 0.06 & 0.03 & 0.01 \\ 
C I     & -0.51 $\pm$ 0.17 & 0.16 & 0.06 & 0.01 & 0.01 \\ 
O I     & -0.14 $\pm$ 0.31 & 0.16 & 0.17 & 0.02 & 0.2 \\ 
Na I    & 0.21 $\pm$ 0.07 & 0.05 & 0.04 & 0.02 & 0.02 \\ 
Mg I    & 0.06 $\pm$ 0.13 & 0.12 & 0.05 & 0.02 & 0.02 \\ 
Al I    & 0.05 $\pm$ 0.29 & 0.16 & 0.15 & 0.04 & 0.19 \\ 
Si II   & 0.08 $\pm$ 0.22 & 0.16 & 0.16 & 0.01 & 0.02 \\ 
Ca II   & -0.1 $\pm$ 0.17 & 0.16 & 0.06 & 0.05 & 0.01 \\ 
Sc II   & -0.41 $\pm$ 0.16 & 0.16 & 0.01 & 0.01 & 0.01 \\ 
Ti II   & 0.5 $\pm$ 0.15 & 0.11 & 0.07 & 0.01 & 0.07 \\ 
Cr II   & 0.39 $\pm$ 0.08 & 0.07 & 0.02 & 0.01 & 0.03 \\ 
Mn I    & 0.19 $\pm$ 0.16 & 0.04 & 0.11 & 0.03 & 0.11 \\ 
Fe I    & 0.37 $\pm$ 0.13 & 0.04 & 0.1 & 0.02 & 0.07 \\ 
Fe II   & 0.34 $\pm$ 0.11 & 0.06 & 0.06 & 0.03 & 0.07 \\ 
Sr II   & 0.92 $\pm$ 0.17 & 0.04 & 0.15 & 0.02 & 0.06 \\ 
Y II    & 0.87 $\pm$ 0.2 & 0.16 & 0.06 & 0.03 & 0.11 \\ 
Ba II   & 1.56 $\pm$ 0.2 & 0.12 & 0.13 & 0.01 & 0.09 \\ 
\hline
\end{tabular}
\normalsize
\label{tab.abunds.BD-08-924}
\end{table}

\begin{table}
\centering
\caption{Chemical abundances for BD-12 905.}
%\hskip -0.10in
\scriptsize
\begin{tabular}{lrcccc}
\hline
\hline
Specie     & [X/H] $\pm$ e$_{tot}$ & e$_{1}$ & e$_{2}$ & e$_{3}$ & e$_{4}$ \\
\hline
Li I    & 2.25 $\pm$ 0.2 & 0.11 & 0.17 & 0.01 & 0.01 \\ 
C I     & -0.18 $\pm$ 0.18 & 0.11 & 0.14 & 0.02 & 0.01 \\ 
O I     & -0.36 $\pm$ 0.12 & 0.02 & 0.11 & 0.02 & 0.05 \\ 
Na I    & 0.1 $\pm$ 0.15 & 0.11 & 0.1 & 0.01 & 0.02 \\ 
Mg I    & 0 $\pm$ 0.17 & 0.06 & 0.16 & 0.02 & 0.01 \\ 
Al I    & -0.71 $\pm$ 0.3 & 0.11 & 0.28 & 0.03 & 0.04 \\ 
Si II   & 0.13 $\pm$ 0.2 & 0.11 & 0.17 & 0.02 & 0.02 \\ 
Ca I    & 0.09 $\pm$ 0.29 & 0.11 & 0.27 & 0.03 & 0.02 \\ 
Ca II   & 0.05 $\pm$ 0.16 & 0.01 & 0.16 & 0.02 & 0.01 \\ 
Sc II   & -0.22 $\pm$ 0.11 & 0.08 & 0.05 & 0.03 & 0.05 \\ 
Ti II   & -0.09 $\pm$ 0.11 & 0.04 & 0.08 & 0.02 & 0.07 \\ 
Cr II   & -0.02 $\pm$ 0.05 & 0.03 & 0.02 & 0.02 & 0.03 \\ 
Mn I    & 0.04 $\pm$ 0.19 & 0.06 & 0.16 & 0.01 & 0.07 \\ 
Fe I    & -0.04 $\pm$ 0.2 & 0.02 & 0.19 & 0.01 & 0.07 \\ 
Fe II   & -0.07 $\pm$ 0.13 & 0.02 & 0.11 & 0.01 & 0.07 \\ 
Co I    & 0.11 $\pm$ 0.14 & 0.11 & 0.09 & 0.02 & 0.02 \\ 
Zn I    & -0.35 $\pm$ 0.12 & 0.02 & 0.11 & 0.01 & 0.04 \\ 
Sr II   & 0.2 $\pm$ 0.2 & 0.04 & 0.19 & 0.01 & 0.04 \\ 
Y II    & -0.1 $\pm$ 0.04 & 0.01 & 0.04 & 0.02 & 0.01 \\ 
Ba II   & 0.31 $\pm$ 0.18 & 0.05 & 0.15 & 0.01 & 0.09 \\ 
\hline
\end{tabular}
\normalsize
\label{tab.abunds.BD-12-905}
\end{table}

\begin{table}
\centering
\caption{Chemical abundances for HD 36726.}
%\hskip -0.10in
\scriptsize
\begin{tabular}{lrcccc}
\hline
\hline
Specie     & [X/H] $\pm$ e$_{tot}$ & e$_{1}$ & e$_{2}$ & e$_{3}$ & e$_{4}$ \\
\hline
C I     & -0.8 $\pm$ 0.18 & 0.13 & 0.12 & 0.01 & 0.01 \\ 
O I     & -0.29 $\pm$ 0.12 & 0.05 & 0.05 & 0.02 & 0.09 \\ 
Mg I    & -1.05 $\pm$ 0.25 & 0.13 & 0.19 & 0.01 & 0.09 \\ 
Mg II   & -1.04 $\pm$ 0.14 & 0.13 & 0.02 & 0.01 & 0.03 \\ 
Si II   & -0.96 $\pm$ 0.16 & 0.13 & 0.08 & 0.01 & 0.01 \\ 
Ca II   & -0.83 $\pm$ 0.26 & 0.16 & 0.2 & 0.01 & 0.05 \\ 
Sc II   & -1.22 $\pm$ 0.19 & 0.13 & 0.14 & 0.01 & 0.01 \\ 
Ti II   & -0.86 $\pm$ 0.13 & 0.06 & 0.12 & 0.01 & 0.03 \\ 
Cr II   & -0.85 $\pm$ 0.05 & 0.01 & 0.05 & 0.01 & 0.01 \\ 
Fe I    & -0.72 $\pm$ 0.2 & 0.03 & 0.2 & 0.01 & 0.03 \\ 
Fe II   & -0.7 $\pm$ 0.13 & 0.04 & 0.11 & 0.01 & 0.03 \\ 
Sr II   & -1.2 $\pm$ 0.24 & 0.03 & 0.23 & 0.01 & 0.04 \\ 
\hline
\end{tabular}
\normalsize
\label{tab.abunds.HD36726}
\end{table}

\begin{table}
\centering
\caption{Chemical abundances for HD 37333.}
%\hskip -0.10in
\scriptsize
\begin{tabular}{lrcccc}
\hline
\hline
Specie     & [X/H] $\pm$ e$_{tot}$ & e$_{1}$ & e$_{2}$ & e$_{3}$ & e$_{4}$ \\
\hline
Mg II   & -0.56 $\pm$ 0.29 & 0.28 & 0.02 & 0.01 & 0.05 \\ 
Si II   & -0.38 $\pm$ 0.28 & 0.28 & 0.04 & 0.01 & 0.01 \\ 
Ca II   & -0.93 $\pm$ 0.33 & 0.26 & 0.18 & 0.02 & 0.08 \\ 
Sc II   & -0.06 $\pm$ 0.29 & 0.26 & 0.1 & 0.01 & 0.09 \\ 
Ti II   & -0.28 $\pm$ 0.11 & 0.07 & 0.08 & 0.01 & 0.01 \\ 
Cr II   & 0.58 $\pm$ 0.09 & 0.03 & 0.07 & 0.01 & 0.04 \\ 
Mn I    & 0.5 $\pm$ 0.25 & 0.1 & 0.23 & 0.02 & 0.01 \\ 
Fe I    & -0.08 $\pm$ 0.2 & 0.05 & 0.19 & 0.02 & 0.03 \\ 
Fe II   & -0.2 $\pm$ 0.14 & 0.1 & 0.07 & 0.02 & 0.06 \\ 
Zn I    & 0.82 $\pm$ 0.28 & 0.26 & 0.11 & 0.02 & 0.01 \\ 
Sr II   & -0.43 $\pm$ 0.35 & 0.26 & 0.22 & 0.01 & 0.07 \\ 
Ba II   & 0.32 $\pm$ 0.32 & 0.26 & 0.18 & 0.01 & 0.01 \\ 
\hline
\end{tabular}
\normalsize
\label{tab.abunds.HD37333}
\end{table}

\begin{table}
\centering
\caption{Chemical abundances for HD 37187.}
%\hskip -0.10in
\scriptsize
\begin{tabular}{lrcccc}
\hline
\hline
Specie     & [X/H] $\pm$ e$_{tot}$ & e$_{1}$ & e$_{2}$ & e$_{3}$ & e$_{4}$ \\
\hline
O I     & -0.2 $\pm$ 0.26 & 0.25 & 0.05 & 0.01 & 0.01 \\ 
Mg I    & 0.17 $\pm$ 0.2 & 0.12 & 0.17 & 0.02 & 0.01 \\ 
Mg II   & -0.1 $\pm$ 0.08 & 0.08 & 0.01 & 0.01 & 0.01 \\ 
Al I    & 0.19 $\pm$ 0.26 & 0.25 & 0.07 & 0.01 & 0.02 \\ 
Si II   & -0.05 $\pm$ 0.09 & 0.07 & 0.04 & 0.02 & 0.01 \\ 
Ca II   & 0.09 $\pm$ 0.3 & 0.25 & 0.16 & 0.02 & 0.02 \\ 
Ti II   & -0.42 $\pm$ 0.13 & 0.06 & 0.11 & 0.01 & 0.01 \\ 
Cr II   & -0.15 $\pm$ 0.07 & 0.06 & 0.04 & 0.01 & 0.01 \\ 
Fe I    & -0.09 $\pm$ 0.15 & 0.08 & 0.13 & 0.01 & 0.03 \\ 
Fe II   & -0.03 $\pm$ 0.09 & 0.07 & 0.06 & 0.01 & 0.02 \\ 
Sr II   & -0.25 $\pm$ 0.3 & 0.25 & 0.16 & 0.01 & 0.01 \\ 
\hline
\end{tabular}
\normalsize
\label{tab.abunds.HD37187}
\end{table}

\begin{table}
\centering
\caption{Chemical abundances for HD 290541.}
%\hskip -0.10in
\scriptsize
\begin{tabular}{lrcccc}
\hline
\hline
Specie     & [X/H] $\pm$ e$_{tot}$ & e$_{1}$ & e$_{2}$ & e$_{3}$ & e$_{4}$ \\
\hline
O I     & -0.71 $\pm$ 0.18 & 0.14 & 0.04 & 0.02 & 0.09 \\ 
Mg I    & -0.32 $\pm$ 0.2 & 0.1 & 0.15 & 0.03 & 0.09 \\ 
Si II   & 0.19 $\pm$ 0.2 & 0.14 & 0.14 & 0.02 & 0.02 \\ 
Ca II   & 0.11 $\pm$ 0.17 & 0.08 & 0.14 & 0.01 & 0.03 \\ 
Sc II   & -0.13 $\pm$ 0.29 & 0.14 & 0.15 & 0.03 & 0.2 \\ 
Ti II   & 0.11 $\pm$ 0.13 & 0.04 & 0.09 & 0.04 & 0.08 \\ 
Cr II   & -0.12 $\pm$ 0.17 & 0.15 & 0.07 & 0.02 & 0.04 \\ 
Mn I    & -0.65 $\pm$ 0.22 & 0.14 & 0.07 & 0.01 & 0.15 \\ 
Fe I    & -0.18 $\pm$ 0.16 & 0.04 & 0.13 & 0.02 & 0.08 \\ 
Fe II   & -0.15 $\pm$ 0.12 & 0.03 & 0.09 & 0.03 & 0.07 \\ 
Ni II   & 0.24 $\pm$ 0.22 & 0.14 & 0.09 & 0.03 & 0.14 \\ 
Sr II   & -0.04 $\pm$ 0.31 & 0.07 & 0.18 & 0.04 & 0.24 \\ 
Y II    & 0.26 $\pm$ 0.16 & 0.14 & 0.04 & 0.03 & 0.04 \\ 
Ba II   & 0.78 $\pm$ 0.28 & 0.25 & 0.1 & 0.01 & 0.09 \\ 
\hline
\end{tabular}
\normalsize
\label{tab.abunds.HD290541}
\end{table}

\begin{table}
\centering
\caption{Chemical abundances for HD 290621.}
%\hskip -0.10in
\scriptsize
\begin{tabular}{lrcccc}
\hline
\hline
Specie     & [X/H] $\pm$ e$_{tot}$ & e$_{1}$ & e$_{2}$ & e$_{3}$ & e$_{4}$ \\
\hline
Li I    & 2.07 $\pm$ 0.21 & 0.15 & 0.14 & 0.01 & 0.01 \\ 
C I     & -0.13 $\pm$ 0.18 & 0.18 & 0.02 & 0.01 & 0.01 \\ 
O I     & -0.21 $\pm$ 0.11 & 0.05 & 0.07 & 0.01 & 0.06 \\ 
Na I    & -0.19 $\pm$ 0.16 & 0.15 & 0.05 & 0.01 & 0.02 \\ 
Mg I    & -0.27 $\pm$ 0.17 & 0.12 & 0.11 & 0.02 & 0.03 \\ 
Al I    & -0.44 $\pm$ 0.29 & 0.15 & 0.24 & 0.02 & 0.05 \\ 
Si II   & -0.03 $\pm$ 0.23 & 0.22 & 0.06 & 0.02 & 0.02 \\ 
Ca I    & -0.04 $\pm$ 0.32 & 0.1 & 0.29 & 0.02 & 0.07 \\ 
Ca II   & 0.08 $\pm$ 0.29 & 0.24 & 0.17 & 0.01 & 0.01 \\ 
Sc II   & -0.28 $\pm$ 0.31 & 0.1 & 0.21 & 0.02 & 0.2 \\ 
Ti II   & -0.03 $\pm$ 0.13 & 0.06 & 0.04 & 0.01 & 0.1 \\ 
Cr II   & -0.05 $\pm$ 0.15 & 0.13 & 0.05 & 0.01 & 0.05 \\ 
Mn I    & 0.18 $\pm$ 0.16 & 0.14 & 0.05 & 0.01 & 0.07 \\ 
Fe I    & -0.12 $\pm$ 0.18 & 0.03 & 0.15 & 0.01 & 0.09 \\ 
Fe II   & -0.13 $\pm$ 0.13 & 0.04 & 0.1 & 0.01 & 0.07 \\ 
Sr II   & 0.17 $\pm$ 0.25 & 0.03 & 0.23 & 0.01 & 0.1 \\ 
Ba II   & 0.18 $\pm$ 0.17 & 0.15 & 0.08 & 0.01 & 0.01 \\ 
\hline
\end{tabular}
\normalsize
\label{tab.abunds.HD290621}
\end{table}

%\FloatBarrier

\section{Average radial velocities and membership}

We present in this section the average radial velocities (obtained from our observations and from literature)
and membership derived in this work (see Table \ref{vr}).
The stars included in the list correspond to the initial membership suggested by \citet{hunt-reffert24}.

\FloatBarrier

\begin{table*}
 \caption{Average radial velocities and membership.} \label{vr}
 \begin{center}
 \scriptsize
 \begin{tabular}{lcccc}
  \hline\hline
  Star      &      $\overline{V}$   &    $\varepsilon_{\overline{V}}$  &     n     &     Cluster \\  
            &        [km s$^{-1}$]  &         [km s$^{-1}$]           &           & \\
  \hline 
HD~28548		            &16.24 & 3.41 & 1 & HSC~1640	 \\  
HD~31624		            &13.48 & 3.67 & 2 & HSC~1640  \\
HD~25674		            &10.73 & 2.36 & 2 & HSC~1640?	 \\
BD-08~924		            &21.15 & 1.48 & 2 & HSC~1640?	 \\
BD-12~905		            &14.21 & 0.56 & 2 & NM	 \\
BD-06~984		            &18.55 & 4.32 & 2 & NM	 \\
TYC~4743-550-1		        &8.68  & 0.79 & 2 & NM	 \\
BD-10~921		            &13.02 & 4.39 & 1 &	HSC~1640 \\
TYC~5317-3258-1 	        &14.17 & 1.68 & 1 &	HSC~1640 \\
TYC~5317-182-1		        &11.61 & 1.68 & 2 &	HSC~1640 \\
TYC~5321-1431-1 	        &-0.51 & 0.83 & 1 & NM	 \\
TYC~5314-2406-1 	        &14.09 & 2.87 & 1 &	HSC~1640 \\
TYC~5322-248-1		        &17.37 & 1.51 & 2 & NM	 \\
UCAC2~26812132 	            &11.94 & 2.03 & 2 &	HSC~1640 \\
UCAC4~421-007258	        &9.71  & 17.25& 1 &	HSC~1640 \\
UCAC4~413-007161	        &15.44 & 6.06 & 1 &	HSC~1640 \\
UCAC4~423-007282	        &11.95 & 2.64 & 1 &	HSC~1640 \\
2MASS~J04283696-1209108     &8.59  & 3.38 & 1 & NM	 \\
PDS~11b		                &49.83 & 10.56& 1 & NM	 \\
UCAC4~398-005446	        &38.25 & 4.65 & 1 & NM \\
UCAC4~402-005416	        &5.5   & 4.78 & 1 & NM	 \\
UCAC4~390-005716	        &9.82  & 4.17 & 1 & NM	 \\
ATO~J065.5183-09.4554	    &4.47  & 11.59& 1 &	HSC~1640 \\
UCAC4~401-006449	        &-10.49& 6.68 & 1 & NM	 \\
UCAC4~401-006561	        &14.55 & 5.42 & 1 &	HSC~1640 \\
UCAC4~398-005446	        &-4.99 & 10.82& 1 & NM	 \\
Gaia~DR3~3188334616743729024&-23.16& 4.95 & 1 & NM	 \\
Gaia~DR3~3185353943800052992&17.98 & 8.97 & 1 &	HSC~1640\\
CRTS~J042853.5-122416	    &63.07 & 7.65 & 1 & NM	 \\
Gaia~DR3~3185262615616971008&20.71 & 8.32 & 1 &	HSC~1640 \\
Gaia~DR3~3179069513293359872&-12.47& 6.21 & 1 & NM	 \\
CRTS~J044139.1-095658	    &37.19 & 6.57 & 1 & NM	 \\
CRTS~J042857.9-150402	    &32.1  & 5.00 & 1 & NM	 \\
Gaia~DR3~3191332503916601728&24.47 & 17.49& 1 &	HSC~1640\\
\hline
HD~37187	           & 16.69 &5.53  & 1 & Theia~139\\
HD~37333	           & 29.13 &1.11  & 2 & NM \\
HD~37845	           & 13.45 &6.08  & 2 & Theia~139\\
HD~36726	           & 17.89 &3.69  & 2 & Theia~139\\
HD~294101A	           & 19.75 &0.84  & 1 & NM \\
HD~290541	           & 13.74 &0.76  & 3 & NM \\
HD~290572 \tablefootmark{a}& 15.79 &1.16  & - & Theia~139\\
HD~290621	           & 15.23 &1.43  & 3 & Theia~139\\
HD~290632	           & 17.55 &1.46  & 2 & Theia~139\\
HD~290548	           & 15.69 &2.63  & 1 & Theia~139\\
HD~294101B	           & 13.75 &13.15 & 1 & Theia~139\\
2MASS~J05281274-0131248& 16.82 &5.36  & 1 & Theia~139\\
2MASS~J05235511-0300024& 14.97 &0.04  & 2 & NM \\
2MASS~J05281224-0131156& 17.8  &13.12 & 1 & Theia~139\\
2MASS~J05453419-0019421& 16.7  &0.02  & 2 & NM \\
2MASS~J05270681-0243277& 15.23 &0.07  & 2 & NM \\
2MASS~J05331323-0215311& 18.79 &3.79  & 1 & Theia~139\\
2MASS~J05374310-0225131& 15.23 &0.02  & 2 & NM \\
ATO~J085.5072-00.4147  & 30.85 &4.27  & 1 & NM \\
**~CVS~65B	           & 26.80 &9.07  & 1 & NM \\
2MASS~J05272021-0239194& -32.37&8.04  & 1 & NM\\
CVSO~175	           & 14.25 &9.66  & 1 & Theia~139\\
\hline  
\end{tabular}
\normalsize
\tablefoot{
Average radial velocities were obtained from our observations and from literature, while membership was derived in this work.
The stars included in the list correspond to the initial membership suggested by \citet{hunt-reffert24}.\\
\tablefoottext{a}{Spectroscopic binary \citep{gaia22}.}
}
\end{center}
\end{table*}

\end{appendix}

\end{document}